
\input harvmac

\Title{\vbox{\baselineskip12pt\hbox{HUTP-92/A051}}}
{\vbox{\centerline{Schwinger-Dyson equation}
       \centerline{in three-dimensional simplicial quantum
                 gravity }}}

\vskip .2cm
\centerline{Hirosi Ooguri}

\vskip .5cm
\centerline{Reserch Institute for Mathematical Sciences,
Kyoto University, Kyoto 606-01, Japan}
\vskip .07cm
\centerline{and}
\vskip .07cm
\centerline{Lyman Laboratory of Physics,
Harvard University, Cambridge, MA 02138, USA}
\vskip .06cm
\centerline{e-mail: ooguri@string.harvard.edu}

\vskip 1.2cm

We study the simplicial quantum gravity in three dimensions.
Motivated by the Boulatov's model which generates a sum
over simplicial complexes weighted with the Turaev-Viro
invariant, we introduce boundary operators in the simplicial
gravity associated to compact orientable surfaces. An amplitude of
the boundary operator is given by a sum over triangulations
in the interior of the boundary surface.
It turns out that the amplitude solves the Schwinger-Dyson equation
even if we restrict the topology in the interior of the surface, as
far as the surface is non-degenerate.
We propose a set of factorization conditions on the amplitudes which
singles out a solution associated to triangulations of $S^3$.

\vskip 1cm
\centerline{\it To be published as an Invited Paper
in Progress of Theoretical Physics}

\Date{September, 1992}


\noindent
{\bf 1. Introduction}

\vskip 0.3cm

In order to understand a quantum theory of gravity,
we need to clarify what we mean by integrations over Riemannian
metrics on a $d$-dimensional differentiable manifold $M$.
One of the possibilities proposed by Regge
\ref\regge{T.Regge, {\sl Nuovo Cimento} {\bf 19} (1961) 558.}
is to discretize the integrals and and replace them by sums over
smooth triangulations on $M$. In this approach, each triangulation $T$
specifies a Riemannian metric on $M$ in the following way.
We impose that an interior of each $d$-simplex in $T$ is
flat Euclidean and that the curvature has a support only on
$(d-2)$-simplexes. Moreover all the edges in $T$ are regarded as being
straight and having the same geodesics length $a$.
(There is another version of the theory where the edges are
allowed to have different lengths \ref\hamber{For a review on recent numerical
results in this approach, see H.Hamber, preprint UCI-Th-91-36 (1991)},
but we do not examine this possibility here.)
These conditions are enough to define a Riemannian metric
associated to the triangulation $T$ of $M$. It is reasonable to expect that
any metric on $M$ can be approximated in this way by using sufficiently
large number of simplexes and by taking the size $a$ of each simplex
to be infinitesimally small.
It is then important to know what types of triangulations
give relevant contributions to the summations. If smooth triangulations
with infinitely many simplexes dominate the summations,
it would make sense to {\it define} the integrals over metrics by the
sums over triangulations.
In two dimensions, this program has worked
rather well and we have learned quite a lot about two-dimensional
quantum gravities from this approach for the last few years.
The purpose of this paper is to study the three-dimensional case.

Let us be more specific about what we mean by the sums over triangulations.
In order to quantize the Einstein gravity, we would like to perform
integrals over metrics $g_{\mu\nu}$
weighted with the exponential of the action
given by $S[g]= \Lambda \int d^d x \sqrt{g} + {1 \over G} \int d^d x
\sqrt{g} R $ where $R$ is the scalar curvature. Suppose that
the metric $g$ is associated to some triangulation $T$ of $M$.
Since each $d$-simplex has the same volume
proportional to $a^d$, the total volume $\int d^d x \sqrt{g}$ of $M$
should be proportional to $a^d n_d(T)$ where $n_i(T)$ ($i=0,1,...,d$)
is the number of $i$-simplexes in $T$. The integral of the scalar
curvature $R$ on the other hand is expressed in the cases
of $d=2$ and $3$ as follows.
When $d=2$, the curvature has a support on the vertices of $T$.
Since the deficit angle around each vertex $v$ is given by
$(2\pi - {\pi \over 3} \tau_\Delta(v))$ where $\tau_\Delta(v)$ is the number
of triangles containing $v$, we obtain
$$ \int d^2 x \sqrt{g} R \sim \sum_{v: vertices}
                         \left( 1- {1 \over 6} \tau_\Delta(v) \right)
                        = n_0(T)-{1 \over 2} n_2(T). $$
Especially when the manifold $M$ is closed, $3n_2(T)=2n_1(T)$ since every
edge on $T$ is shared by exactly two triangles. In this case, the
right-hand side in the above becomes $\chi
= n_0(T)-n_1(T)+n_2(T)$, the Euler number of $M$.
When $d=3$, the curvature is concentrated on the edges of $T$
and is characterized by the deficit angle around
each edge $e$ given by $(2\pi - \cos^{-1}({1 \over 3}) \tau_t(e))$
where $\tau_t(e)$ denotes the number of tetrahedra containing $e$.
Since each edge has the geodesic length $a$, the integral of
the scalar curvature is given by
$$ \eqalign{ &\int d^3 x  \sqrt{g} R  \sim a \sum_{e: edges}
                   \left[ 2 \pi -
         \cos^{-1} \left( {1 \over 3} \right)
        \tau_t(e) \right]
    \cr &    = a \left[ 2 \pi n_1(T) -
         6 \cos^{-1} \left( {1 \over 3} \right) n_3(T)
                         \right] . \cr}$$
Therefore, for both $d=2$ and $3$, the action $S[g]$ is
expressed as a linear combination of $n_d(T)$ and $n_{d-2}(T)$.
$$  S[g] = - \kappa n_{d-2}(T) + \beta n_d(T) . $$

The partition function of the simplicial quantum gravity is then
defined by
\eqn\partition{ Z_d = \sum_{T} {1 \over C(T)} e^{\kappa n_{d-2}(T)
              - \beta n_d(T)}  ,}
where the sum is over all possible triangulations of
$M$, and the factor $C(T)$ is the order of
a symmetry of $T$ if any. This factor is included in order
to take into account the volume factor of the diffeomorphism group.
First of all, we need to know if this summation is convergent at all.
Since $Z_d$ can be written as
$$ Z_d = \sum_{n_d,n_{d-2}=0}^\infty Z_{n_d,n_{d-2}}(M)
                       e^{\kappa n_{d-2} - \beta n_d} $$
where $Z_{n_d,n_{d-2}}(M)$ is a number of triangulations of $M$
with given values of $n_d$ and $n_{d-2}$ weighted with the
factor $C(T)$, we would like to know
the asymptotic behavior of $Z_{n_d,n_{d-2}}(M)$
for large values of $n_d$ and $n_{d-2}$.
In two dimensions, $n_2$ and $n_0$ are not independent and they are
related as
as $n_0={1 \over 2}n_2+\chi$. Especially when $\chi=2$ (i.e.
$M \simeq S^2$), the exact asymptotic behavior of $Z_{n_2,n_0}$
is known from the combinatorial study of ref. \ref\tutte{W.T.Tutte,
{\sl Can. J. Math.} {\bf 14} (1962) 21.} as
\eqn\twosshere{
     \eqalign{ Z_{n_2,n_0}(S^2) & \sim n_2^{-{7 \over 2}}
                      e^{\beta_c n_2}
        ~~~~~~(n_2 \rightarrow \infty,~n_0={1 \over 2}n_2+2) \cr
          &~~~~
    e^{\beta_c} = { 2^4 \over 3^{3/2}} . \cr} }
The sum over triangulations can also be evaluated by the matrix model
\ref\bipz{E.Br\'ezin,
C.Itzykson, G.Parisi and J.-B.Zuber, {\sl Commun. Math. Phys.}
{\bf 59} (1978) 35.}, and the above formula for $S^2$ is
reproduced by
the sum over one-particle irreducible vacuum diagrams \bipz\
in the large-$N$ limit of the matrix model
\ref\eguchi{T.Eguchi and H.Kawai, {\sl Phys. Lett.}
{\bf 114B} (1982) 247.} \ref\david{F.David, {\sl Nucl. Phys.}
{\bf B257} (1985) 45} \ref\kazakov{V.A. Kazakov, {\sl Phys. Lett.}
{\bf 150B} (1985) 282.}.
(The sum over connected vacuum diagrams
also gives a similar formula with a different value of $\beta_c$;
$e^{\beta_c} = 2 \cdot 3^{3\over 4}$.)
When $M$ is a surface of higher genus ($\chi \leq 0$),
subleading corrections \ref\bessis{D.Bessis, C.Itzykson
and J.-B.Zuber, {\sl Adv. Appl. Math.} {\bf1} (1980) 109 .}
to the large-$N$ matrix model give \ref\metha{I.Kostov and M.L.Metha,
{\sl Phys. Lett.} {\bf 189B} (1987) 118.}
$$ Z_{n_2 n_0}(M) \sim
            n_2^{-{5 \over 4}\chi-1}
               e^{\beta_c n_2}~~~~~
     (n_2 \rightarrow \infty, ~n_0={1\over 2} n_2 + \chi). $$
Therefore the contribution to $Z_{d=2}$
from $T$ with many triangles is expressed as
$$  Z_{d=2} \sim e^{\kappa \chi}
            \sum_{n_2} n_2^{-{5 \over 4} \chi-1}
              e^{-(\tilde{\beta}-\beta_c)n_2}
 \sim \cases{ (\tilde{\beta} - \beta_c)^{{5 \over 4}\chi}
                    & if $\chi \neq 0$; \cr
       \log \left( {1 \over \tilde{\beta} - \beta_c} \right)
                    & if $\chi = 0$    \cr }  $$
where $\tilde{\beta}=\beta-{1\over 2}\kappa$, and it is
convergent for $\tilde{\beta} > \beta_c$.  Moreover as $\tilde{\beta}$
approaches to the critical value $\beta_c$, the contributions from
triangulations with $n_2>>1$ become more and more significant for $Z_{d=2}$
with $\chi \leq 0$ and for $\partial^3_\beta Z_{d=2}$ with $\chi = 2$.
Thus we have a good reason to believe that the program of the simplicial
quantum gravity works well in two dimensions.  Indeed this approach has
turned out to be quite successful and its validity has been confirmed by
comparisons of the matrix model computations with the results of the
continuum field theories \ref\kpz{
V.G.Knizhnik, A.M.Polyakov and A.B.Zamolodchikov, {\sl
Mod. Phys. Lett.} {\bf A3} (1988) 819.} \ref\ddk{F.David, {\sl Mod. Phys.
Lett.} {\bf A3} (1988) 1651; J.Distler and H.Kawai, {\sl Nucl. Phys} {\bf
B321} (1989) 509.}
\ref\double{M.Douglas and S.Shenker, {\sl Nucl. Phys.} {\bf B335} (1990)
635; D.J.Gross and A.A.Migdal, {\sl Phys. Rev. Lett.} {\bf 64}
(1990) 127; E.Brezin and V.A.Kazakov, {\sl Phys. Lett.} {\bf B236}
(1990) 144.}
\ref\witten{E.Witten, {\sl Nucl. Phys.} {\bf B340} (1990) 281;
R.Dijkgraaf and E.Witten, {\sl Nucl. Phys.} {\bf B342} (1990) 342.}
\ref\dvv{R.Dijkgraaf, E.Verlinde and H.Verlinde, {\sl Nucl. Phys.}
{\bf B348} (1991) 435; E.Verlinde and H.Verlinde, {\sl Nucl. Phys.}
{\bf B348} (1991) 457.}
\ref\fkn{M.Fukuma, H.Kawai and R.Nakayama, {\sl Int. J. Mod. Phys.}
{\bf A6} (1991) 1385.}
\ref\moore{G.Moore, N.Seiberg and M.Staudacher, {\sl Nucl. Phys.} {\bf B362}
(1991) 665; G.Moore and N.Seiberg, {\sl Int. J. Mod. Phys.}
{\bf A7} (1991) 190.}.

Recently several numerical results have been reported on the
three-dimensional simplicial quantum gravity \ref\ambjorn{
J.Ambj\o rn and S.Varsted, {\sl Phys. Lett.} {\bf 266B} (1991)
285.} \ref\migdal{
M.E.Agishtein and A.A.Migdal, {\sl Mod. Phys. Lett.} {\bf A6}
(1991) 1863.} \ref\denmark{D.V.Boulatov and A.Krywicki
{\sl Mod. Phys. Lett.} {\bf A6} (1991) 3005.}
\ref\denmarktwo{ J.Ambj\o rn and
S.Varsted, preprint NBI-HE-91-45 (1991)} \ref\denmarkthree{
J.Ambj\o rn,
D.V.Boulatov, A.Krywicki and S.Varsted, preprint NBI-HE-91-46
(1991).}. Their results seem to be compatible with
the following bound on the number of triangulations
of $S^3$.
\eqn\bound{  Z_{n_3}(S^3)=\sum_{n_1} Z_{n_3 n_1}(S^3)
         < C e^{\beta_c n_3} ~~~~~
             (n_3 \rightarrow \infty ), }
for some constants $C$ and $\beta_c$.
In three dimensions, the Euler number $\chi=n_0-n_1+n_2-n_3$
is zero for any triangulation of a closed manifold.
We also have
$n_2=2n_3$ which follows from the fact that any triangle in
$T$ is shared by exactly two tetrahedra. Therefore
$n_3 = n_1-n_0$, and $n_3$ should always be less than $n_1$.
On the other hand, since
each edge in $T$ belongs to at least one tetrahedra,
$n_1$ must be less than or equal to $6n_3$. Namely
$n_3 < n_1 \leq 6n_3$. Thus, if the exponential bound
\bound\ is indeed correct, the sum over triangulations in
eq. \partition\ would be  bounded as
$$ Z_{d=3} \leq
           \sum_{n_3,n_1} Z_{n_3 n_1} e^{-(\beta-6\kappa)n_3}
         < C \sum_{n_3} e^{-(\beta-6\kappa-\beta_c)n_3} $$
for $\kappa \geq 0$ and
$$ Z_{d=3} < \sum_{n_3,n_1} Z_{n_3 n_1} e^{-(\beta-\kappa)n_3}
           < C \sum_{n_3} e^{-(\beta-\kappa -\beta_c)} $$
for $\kappa < 0$.
The sum \partition\ would then be convergent at least for
$\beta-6\kappa > \beta_c$ ($\kappa \geq 0$) and
for $\beta-\kappa >\beta_c$ ($0 > \kappa$).
It is thus important to develop an analytical tool to examine
the behavior of $Z_{n_3}(S^3)$.

In two dimensions, the matrix model \bipz\ has been found
useful to study the simplicial quantum gravity, and there have
been some attempts to extend it to three dimensions \ref\tensor{
J.Ambj\o rn, B.Durhuus and T.Jonsson, {\sl Mod. Phys. Lett.}
{\bf A6} (1991) 1133.} \ref\sasakura{N. Sasakura,
{\sl Mod. Phys. Lett.} {\bf A6} (1991) 2613.} \ref\gross{
M.Godfrey and M.Gross, {\sl Phys. Rev.} {\bf D43} (1992) R1749.}.
In this so-called tensor model, one considers and integral over
rank-three tensors rather than matrices, and the perturbative
expansion of the integral generates simplicial complexes, i.e.
collections of tetrahedra
whose boundary triangles are pairwisely identified. However
the collections of tetrahedra generated in this model
are not necessarily triangulations of smooth manifolds.
In fact their Euler numbers are not always zero, but in general
non-negative integers \tensor .

The basic reason for the failure of the tensor model is that the model does
not contain as many parameters as one needs to distinguish various
topologies in three dimensions.  In two dimensions, a topology of
a closed orientable surface is characterized by the number of handles, and
two parameters in the model are enough to control both the topology and the
volume $n_2$ of the surface. The matrix model indeed has two parameters,
the size $N$ of the matrix and the coupling constant $g$.  In three
dimensions, on the other hand, we need at least three parameters to control
three independent variables $n_3$, $n_1$ and $\chi=n_0-n_1+n_2-n_3$.  The
simplest version of the tensor model has only two parameters, the size $N$
of the tensor and the coupling constant, and it is not possible to force
$\chi$ to be zero.

Recently, Boulatov proposed an improved version of the tensor model
\ref\boulatov{D.V.Boulatov, preprint SPhT/92-017 (1992).} by
incorporating a gauge symmetry of a finite group or a quantum group.
If one takes the gauge group to be a finite group $G$, the order of $G$
couples to the Euler number of the simplicial complex.
For example, when $G$ is the cyclic group $Z_p$, the partition function
of the model is expressed as a sum over simplicial complexes $T$
weighted with a factor $p^{\chi(T)}|H_1(T,Z_p)|$, where $|H_1(T;Z_p)|$
is the order of the first cohomology group of $T$
with coefficients in $Z_p$.
In three dimensions, the necessary and the sufficient condition
for a simplicial complex $T$ to be a manifold is $\chi(T) =0$,
and otherwise $\chi(T) > 0$.
Thus, if one could make sense of the limit of $p \rightarrow 0$,
the perturbative expansion of the model in this limit
would be dominated by the sum over simplicial manifolds with $\chi(T)=0$.

The $p \rightarrow 0$ limit of the $Z_p$ model not only imposes
$\chi=0$ but gives an additional condition on topologies of simplicial
manifolds, i.e. the first and the second homologies of the manifolds
are restricted to be trivial \boulatov . Without any condition
on the topology, the number of three-dimensional simplicial manifolds
with a given number $n_3$ of tetrahedra grows at least factorially
in $n_3$ \tensor , and the sum over simplicial manifolds becomes divergent.
Thus one needs to impose restrictions on the topologies.
It is not clear if the conditions on the homologies are enough
to tame the factorial growth since there are many topologies
in three dimensions with the same homologies as those of
$S^3$. If not,
the sum over triangulations generated by the $Z_p$ model would still be
divergent, and we would need
to develop a mechanism to impose further restrictions on the topologies.

In this paper, we examine the Boulatov's model associated to the quantum
group $SU_q(2)$. In this model, the sum over simplicial complexes
is weighted with $\Lambda_q^{\chi(T)} I_q(T)$ where
$$ \Lambda_q = - {2(k+2) \over (q^{1/2}-q^{-1/2})^2}~,~~
              q=e^{2\pi i/(k+2)} $$
and $I_q(T)$ is the Turaev-Viro invariant  \ref\viro{V.G. Turaev
and O.Y. Viro, ``{\it State Sum Invariants of $3$-Manifolds
and Quantum $6j$-Symbols}'' (1990).}.
We define boundary operators of the model associated to
compact orientable triangulated two-dimensional surfaces, and derive the
Schwinger-Dyson equation for the amplitudes of the operators. It turns
out that the Schwinger-Dyson equation takes the same form for any value
of $q$ when the surfaces involved are non-degenerate.
Since the amplitude of the boundary operator in the Boulatov model
is expressed a sum over triangulations in the interior
of the boundary surface weighted with $\Lambda_q^{\chi} I_q$
depending on $q$, the independence of the Schwinger-Dyson equation
on $q$ suggests that we can restrict the topology of $T$ in the interior
of the surface without spoiling the equation. Indeed
we find that, associated to an arbitrary closed orientable three-dimensional
manifold, one can construct a solution to the equation.
In order to single out a solution associated
to triangulations of $S^3$, we introduce a set of factorization
conditions for degenerate surfaces with which the equation can be
solved uniquely and inductively.

In the case of the two-dimensional simplicial gravity,
the Schwinger-Dyson equation can be derived from purely
combinatorial considerations without using the matrix model.
This is also true in the three-dimensional model studied here.
However, in order
to take into account the symmetry factor to each graph,
it is more convenient and transparent to employ the
perturbative expansion of the Boulatov model.

This paper is organized as follows. In Section 2, we briefly describe
of the Schwinger-Dyson equation in the large-$N$ matrix model and its
combinatorial meaning. We then define the Boulatov's model in
Section 3 and examine its properties. The boundary operator are
defined in Section 4, and the Schwinger-Dyson equation is derived.
In Section 5, we show that there is a solution to the equation
associated to each closed orientable three-dimensional manifold.
The factorization conditions on the amplitude
which characterize the solution associated to $S^3$ are
defined in Section 6.
The Schwinger-Dyson equation combined with the
factorization conditions is shown to have a unique solution.

\vskip 1cm

\noindent
{\bf 2. Schwinger-Dyson Equation in the Matrix Model}

\vskip 0.3cm

Let $(M_{ij})$ be a $N \times N$ hermitian matrix and consider the
following integral.
\eqn\mat{ Z = \int [ dM ] \exp(-S(M)) }
where
$$ S(M) = {1 \over 2} {\sl tr} M^2 - {g \over 3 \sqrt{N}}
                                          {\sl tr} M^3 . $$
%
by expanding the integral
\mat\ in powers of $g$, $Z$ is expressed as a sum over orientable
closed trivalent graphs, each of which is dual to a triangulation
of an orientable closed (but not necessarily connected) surface.
A sum over closed and connected simplicial manifolds
is obtained by taking the logarithm of $Z$ as
$$ \log Z = \sum_{T} {1 \over C(T)}
       \left( {g \over \sqrt{N}} \right)^{n_2(T)}
                         N^{n_0(T)}
     = \sum_{T} {1 \over C(T)} g^{n_2(T)} N^{\chi(T)} . $$
Especially the sum over triangulations of $S^2$ is given
by the leading term in the large-$N$ limit of $\log Z$.

The basic quantity we examine in this section is
the expectation value of ${\sl tr} M^n$ at the large-$N$.
$$ u_n = \lim_{N \rightarrow \infty}
      {1 \over N^{{1\over 2}n +1} } \langle {\sl tr} M^n \rangle
               = \lim_{N \rightarrow \infty}
     {1 \over N^{{1 \over 2}n+1}} \cdot {1 \over Z}
                   \int [d M]
            {\sl tr} M^n    \exp (-S(M)). $$
The power of $N$ is multiplied to $\langle {\sl tr} M^n \rangle$
so that $u_n$ gives a finite value in the limit of $N \rightarrow \infty$.
The Schwinger-Dyson equation for $u_n$
\ref\wadia{S.R.Wadia, {\sl Phys. Rev.} {\bf D24} (1981) 970.},
\ref\migdalpr{A.A.Migdal, {\sl Phys. Rep.} {\bf 102} (1983) 199.}
is derived from
\eqn\sdtwo{\eqalign{0&={1 \over Z}
             \int [dM] {\partial \over \partial M_{ij}}
             \left[ (M^{n-1})_{ij} \exp (-S(M)) \right] \cr
         &= -\langle {\sl tr} M^{n} \rangle
            +{g \over \sqrt{N}}
                  \langle {\sl tr} M^{n+1} \rangle
              + \sum_{k=0}^{n-2} \langle {\sl tr} M^{k}
                           {\sl tr} M^{n-k-2} \rangle . \cr} }
Here the integration by part is justified in each term in
the perturbative expansion in $g$.
In the limit of large-$N$,
the last term in the right-hand side of eq. \sdtwo\ factorizes as
$$ {1 \over N^{{1 \over 2}k+1}}{1 \over N^{{1\over 2}(n-k-2)+1} }
         \langle  {\sl tr} M^{k}   {\sl tr} M^{n-k-2} \rangle
     = u_{k} u_{n-k-2} + O({1 \over N^2}) $$
since only the planar graphs contribute.
Thus we obtain a recurrence relation for $u_n$ as
\eqn\rec{ u_{n} = g u_{n+1} + \sum_{k=0}^{n-2} u_{k} u_{n-k-2} .}
This is the Schwinger-Dyson equation in the large-$N$ limit of
the matrix model.

This equation has the following simple graphical interpretation.
The expectation value $u_n$ can be expressed as a sum over
planar trivalent graphs with $n$ external legs. The graphs
are not necessarily connected, but they do not contain
a closed disconnected part because of the factor $1/Z$ in $u_n$.
Let us take one of the graphs $T$ in the summation and pay attention
to one of the external legs on it.
The leg should be either connected to a trivalent vertex
in $T$ or to another external leg.
In the former case, the rest will be a graph
with $(n+1)$ external legs. If on the other hand the leg is
connected to another external legs, the line connecting
these two legs will separate
the original planar graph into two disconnected graphs
with $k$ and $(n-k-2)$ external legs ($k=1,2,...,n-2$). The two terms
in the right-hand side of the Schwinger-Dyson equation
correspond to these two possibilities.

It is also straightforward to show that the Schwinger-Dyson equation
completely determines the power series expansion
of $u_n$ in $g$. By substituting
$$ u_n(g) = \sum_{n_2=0}^\infty u_{n,n_2} g^{n_2} $$
into eq. \rec , we obtain a recurrence relation with respect to
$n$ and $n_2$ as
\eqn\biblioord{ u_{n,n_2} = u_{n+1,n_2-1} + \sum_{k=0}^{n-2}
      \sum_{s=0}^{n_2} u_{k,s} u_{n-k-2,n_2-s}. }
Since $u_0(g)$ is normalized to be $1$, the initial
condition of the recurrence relation is $u_{0,n_2}=\delta_{0,n_2}$,
with which $u_{n,n_2}$ is determined inductively by solving
\biblioord\ following the bibliographic ordering:
$(n,n_2) \succ (n',n_2')$ if $n_2>n_2'$ or $n_2=n_2'$, $n>n'$.

Since the nonlinear term in the Schwinger-Dyson equation \rec\ is the
convolution of $u_n$'s, it can be transformed into the integral
equation
\eqn\integral{ -1 +g \lambda^2 = {\sl P}\int {d \lambda' \over 2 \pi}
               {\rho(\lambda') \over \lambda - \lambda'} }
where
\eqn\laplace{ u_n = \int d \lambda \rho(\lambda) \lambda^n }
and
${\sl P}\int$ denotes the principal part of the integral.  This equation
can be solved using the technique of the Riemann-Hilbert correspondence
\bipz\ as
$$ \eqalign{ \rho(\lambda) &= {1 \over \pi}[1+{1\over 2}
                             g(a+b)+g\lambda]
        \sqrt{(\lambda-a)(b-\lambda)} ~~(a \leq \lambda \leq b) \cr
       &= ~~ 0 ~~~~~~~~~~~~~~~~~~~~~(\lambda < a~~or~~b < \lambda) \cr}$$
where $a$ and $b$ are determined by
$$ g(b-a)^2 + 2(a+b)[ 2 +g(a+b)] = 0~,~~(b-a)^2[1+g(a+b)]=1. $$

We can also compute
the partition function \mat\ itself from the solution to
the Schwinger-Dyson equation. By using the relation
between $u_3$ and $Z$,
\eqn\matrixdel{  u_3 = {1 \over N^2 \sqrt{N}}
           \langle {\sl tr} M^3 \rangle
= {3 \over N^2} {1\over Z}{\partial Z \over \partial g}  }
we obtain
$$ Z(g) = Z(0) \exp({N^2\over 3} \int_0^g dg'  u_3(g')). $$
The approach described here using the Schwinger-Dyson equation is
in fact equivalent at the large-$N$ to the standard computation
\bipz\ \bessis\ using the eigenvalues $\lambda_i$ ($i=1,...,N$)
of $M$. By inverting the integral transformation \laplace ,
we can express $\rho(\lambda)$ in terms of $u_n$ as
$$ \eqalign{ \rho(\lambda) &
          = \int {dk \over 2 \pi} e^{-ik\lambda}
                            \sum_{n=0}^\infty {1 \over n!}
                                  u_n \lambda^n
            = \int {dk \over 2 \pi}
    e^{-ik\lambda} \langle {1 \over N} {\sl tr}
           \exp( ik { M \over \sqrt{N} } ) \rangle  \cr
            & = \sum_{i=1}^N {1 \over N} ~\langle
             \delta( \lambda - {\lambda_i \over \sqrt{N}} )\rangle . \cr} $$
Namely $\rho(\lambda)$ is the density of the eigenvalues
of $M$.

%
%

\vskip 1cm

\noindent
{\bf 3. Boulatov's Model}

\vskip 0.3cm
In this section, we describe the model due to Boulatov \boulatov\
and examine its properties. Let us consider the following integral
\eqn\boulatovmodel{ Z_q = \int [dM] \exp(-S(M))  }
with $S(M)$ given by
\eqn\boulatovaction{
     \eqalign{ & S(M) = {1 \over 6} \sum_{a_1,a_2,a_3=1}^{N}
                      \sum_{ \{ j_1,j_2,j_3 \} }
                      \sum_{\{ -j_k \leq m_k \leq j_k \}}
                    |M^{j_1j_2j_3;a_1a_2a_3}_{m_1m_2m_3}|^2 \cr
       &~~-{g \Lambda_q \over 12}
        \sum_{a_1,...,a_6=1}^N  \sum_{ \{ j_1,...,j_6\} }
           \sum_{ \{ -j_k \leq m_k \leq j_k \} } (-1)^{\sum_{k=1}^6j_k}
           \left\{ \matrix{ j_1 & j_2 & j_3 \cr
                            j_4 & j_5 & j_6 \cr } \right\}_q  \cr
     &~~~~~~\times
        (-q)^{\sum_{k=1}^6 m_k}  M^{j_1j_2j_3;a_1a_2a_3}_{-m_1-m_2-m_3}
          M^{j_3j_4j_5;a_3a_4a_5}_{m_3-m_4m_5}
         M^{j_1j_5j_6;a_1a_5a_6}_{m_1-m_5m_6}
          M^{j_2j_6j_4;a_2a_6a_4}_{m_2-m_6m_4}  \cr  } }
where
$$  \left\{ \matrix{ j_1 & j_2 & j_3 \cr
                     j_4 & j_5 & j_6 \cr } \right\}_q $$
is the Racah-Wigner $6j$-symbol of the quantum group $SU_q(2)$
\ref\resh{A.N.Kirillov and N.Yu.Reshetikhin, in {\it Infinite
Dimensional Lie Algebras and Groups,} edited by V.G. Kac
(World Scientific, 1989). } with $q = e^{2 \pi i/(k+2)}$
and
$$ \Lambda_q = -{ 2 (k+2) \over (q^{1/2} - q^{-1/2})^2 } .$$
The integration variable $M^{j_1j_2j_3;a_1a_2a_3}_{m_1m_2m_3}$
obeys the conditions
$$ \eqalign{
 & (1)~ M_{j_1j_2j_3;a_1a_2a_3}^{m_1m_2m_3}=0 ~~~ {\sl unless} ~
  |j_1-j_2| \leq j_3 \leq j_1+j_2 ~ {\sl and} ~ j_1+j_2+j_3 \leq k  \cr
 & (2)~ \overline{M}^{j_1j_2j_3;a_1a_2a_3}_{m_1m_2m_3} =
     (-1)^{j_1+j_2+j_3}(-q)^{m_1+m_2+m_3}
           M^{j_1j_2j_3;a_1a_2a_3}_{-m_1-m_2m_3}  \cr
 & (3)~ M^{j_1j_2j_3;a_1a_2a_3}_{m_1m_2m_3} =
         M^{j_2j_3j_1;a_2a_3a_1}_{m_2m_3m_1} . \cr} $$
The perturbative expansion of this integral
gives a sum over Feynman graphs each of which is
dual to an oriented simplicial complex in three dimensions.
A contribution of each simplicial complex $T$
is evaluated by assigning a triplet of indices
$(a_e,j_e,m_e)$ to each edge $e$ in $T$ and by summing over the indices
on all the edges with the conditions
$$     |j_{e_1(\Delta)} - j_{e_2(\Delta)}|  \leq j_{e_3(\Delta)}
               \leq j_{e_1(\Delta)} + j_{e_2(\Delta)}~,~~
            j_{e_1(\Delta)}+j_{e_2(\Delta)}+j_{e_3(\Delta)} \leq
                       k   $$
for every triangle $\Delta$ in $T$ where $e_r(\Delta)$
($r=1,2,3$) are the edges of $\Delta$.
We then obtain
$$ \log Z_q = \sum_T {1 \over C(T)}(g\Lambda_q)^{n_3(T)} N^{n_1(T)}
      \Lambda_q^{n_0(T)-1} I_{q}(T) ,$$
where the sum $\sum_T$ is over connected simplicial complexes,
and $I_{q}(T)$ is given by
\eqn\topological{ \eqalign{ I_{q}(T) = & {1 \over \Lambda_p^{n_0(T)-1}}
    \sum_{\{ j \} } \prod_{e:edges} (-1)^{2j_e}[ 2j_e + 1 ]_q \cr
     &~~\times \prod_{t:tetrahedra} (-1)^{\sum_{k=1}^6 j_{e_k(t)}}
 \left\{ \matrix{ j_{e_1(t)} & j_{e_2(t)} & j_{e_3(t)} \cr
                  j_{e_4(t)} & j_{e_5(t)} & j_{e_6(t)} \cr }
             \right\}_q      \cr}  }
where
$$ [ 2j+1 ]_q = { q^{j+1/2} - q^{-j-1/2} \over
                    q^{1/2} - q^{-1/2} }, $$
and $e_r(t)$ ($r=1,...,6$) are the six edges of the tetrahedron $t$.

It has been shown by Turaev and Viro \viro\
that $I_q(T)$ is a topological invariant
of $T$, i.e. $I_q(T_1)= I_q(T_2)$ if
$T_1$ and $T_2$ are combinatorially equivalent. Especially when
$T$ is a triangulation of an orientable manifold $M$, it has been shown in
\ref\turaev{V.G. Turaev, {\sl C. R. Acad. Sci. Paris, t. 313, S\'erie
I} (1991) 395; "{\it Topology of Shadow,}" preprint (1991).}
\ref\walker{
K. Walker, "{\it On Witten's $3$-Manifold
Invariants,}" preprint (1990).} that $I_q(T)$ is equal to $|\tau_q(M)|^2$
where $\tau_q(M)$ is the Witten-Reshetikhin-Turaev invariant
\ref\witten{E. Witten,
{\sl Commun. Math. Phys.} {\bf 121} (1989) 351;
N.Y.Reshetikhin and V.G.Turaev, {\sl Invent. Math.} {\bf 103}
(1991) 547.}. (The relation
between $I_q(T)$ and $\tau_q(M)$ at $q \rightarrow 1$
has also been studied in \ref\sasakura{H. Ooguri and N.Sasakura,
{\sl Mod. Phys. Lett.} {\bf A6} (1991) 3591.}
\ref\ooguri{H. Ooguri,  {\sl Nucl. Phys.} {\bf B382}
(1992) 276.}. See also \ref\williams{
F.Archer and R.Williams, {\sl Phys. Lett.} {\bf B273} (1991) 438.}
and \ref\mizo{S.Mizoguchi and T.Tada, {\sl Phys. Rev. Lett.}
{\bf 68} (1992) 1795.}.)
For example
$$ I_q(S^3) = 1 ~,~~~I_q(S^1 \times S^2)
                    = \Lambda_q . $$
In general, it has been shown by Kohno \ref\kohno{T.Kohno,
{\sl Topology} {\bf 31} (1992) 203.} that
$I_q(M)$ is bounded from above as
\eqn\genus{ I_q(M) \leq \Lambda_q^{h(M)}  }
where $h(M)$ is the Heegaard genus. Any three-dimensional manifold
can be decomposed into a pair of handlebodies ${\cal H}_h$ of genus $h$
glued together on their boundaries (the Heegaard splitting),
and the Heegaard genus $h(M)$ is the minimum number of such $h$.
Especially $h(M)=0$ if and only if $M$ is homeomorphic to
$S^3$ and otherwise $h(M) \geq 1$.

By introducing a new variable $\mu$ by $N=\mu/\Lambda_q$,
the partition function $Z_q$ is expressed as
\eqn\tensorpartition{
    \log Z_q = {1 \over \Lambda_q}
                  \sum_T
    {1 \over C(T)}     g^{n_3(T)} \mu^{n_1(T)}
    \Lambda_q^{\chi(T)} I_q(T) . }
Suppose that we could take the limit $q \rightarrow 0$
so that $\Lambda_q \rightarrow 0$
while maintaining the inequality \genus .
As we mentioned in the introduction,
$\chi(T) \geq 0$ for any simplicial complex $T$ in three dimensions
and the equality holds if and only if $T$ is a simplicial manifold.
Because of the inequality \genus,
the $q \rightarrow 0$ limit of $\log Z_q$ would be dominated
by a sum over simplicial manifolds with $\chi(T)=0$ and
$h(T)=0$, i.e. $T \simeq S^3$.  Unfortunately, it is not clear
how to make sense of this limit since the model is defined only
when $q$ is a root of unity. In the following,
we will develop an alternative mechanism
to impose the condition $T \simeq S^3$ without using the
$q \rightarrow 0$ limit.

\vskip 1cm

\noindent
{\bf 4. Schwinger-Dyson Equation in Three Dimensions}

\vskip 0.3cm

In the case of the matrix model, the integral over the
hermitian matrix $M$ in \partition\ can be reduced to
an integral over the eigenvalues
$\lambda_i$ ($i=1,...,N$) of $M$
$$ \int [d M] \exp(-S(M)) = \int \prod_{i=1}^N d \lambda_i
                     \prod_{i<j}(\lambda_i-\lambda_j)^2
        \exp(-\sum_i ({1 \over 2}\lambda_i^2-{g \over 3 \sqrt{N}}
                         \lambda_i^3) ), $$
and the powerful methods of the WKB approximation \bipz\
and the orthogonal polynomials \bessis\ have been used
to successfully analyze the model. This approach, however,
does not seem to have an obvious generalization to three dimensions.
It is not clear if the integral over $M^{j_1j_2j_3;a_1a_2a_3}_{m_1m_2m_3}$
could be significantly reduced by introducing some variables similar to
the eigenvalues.

In Section 2, we saw that the large-$N$ matrix model
can also be studied by using the Schwinger-Dyson equation
for $u_n$, the expectation value of ${\sl tr} M^n$. In fact
$u_n$ is the integral transform \laplace\
of $\rho(\lambda)$, the density of the eigenvalue. This is not too surprizing
since both the WKB approximation and the Schwinger-Dyson equation
at large-$N$ make use of the classical equation of motion of the model.
The point here is that the method of the Schwinger-Dyson equation
is applicable to the three-dimensional case.

In case of the matrix model, the expectation value of ${\sl tr} M^n$
is expressed as a sum over planar graphs with $n$ external legs.
Thus ${\sl tr} M^n$ may be regarded as an operator which
creates a circle with $n$ edges in each of the planar graphs.
The natural thing to
consider in three dimensions should then be an operator which creates
a closed connected triangulated surface $\Sigma$. Indeed we can
construct such an operator as follows. We attach a triplet of indices
$(j_e,m_e,a_e)$ to each edge $e$ on the surface $\Sigma$
and associate
$M^{j_1j_2j_3;a_1a_2a_3}_{m_1m_2m_3}$
($j_r = j_{e_r(\Delta)}$, $m_r = \pm m_{e_r(\Delta)}$ ,
$a_r = a_{e_r(\Delta)}$) to each triangle $\Delta$.
The signs $(\pm)$ of $m_{r}$ for a pair of triangle sharing
the edge $e$ are chosen to be opposite. We then define
a polynomial $F_\Sigma( j ; M)$ of $M$
by summing over $\{ m \}$ and $\{ a \}$ as
\eqn\surfacepoly{ \eqalign{ F_\Sigma( j  ; M) =&
   \sum_{ \{ m \} , \{ a \} } \prod_{e:edges} {(-1)^{j_e} \over 2}
               \left[ (-q)^{m_e}+(-q)^{-m_e} \right] \cr
                &~~~\times    \prod_{\Delta: triangles}
              M^{j_{e_1(\Delta)}j_{e_2(\Delta)}j_{e_3(\Delta)}
                 a_{e_1(\Delta)}a_{e_2(\Delta)}a_{e_3(\Delta)}}
                    _{\pm m_{e_1(\Delta)}
               \pm m_{e_2(\Delta)} \pm m_{e_3(\Delta)}} .\cr} }
By taking an expectation value of $F_\Sigma$, we obtain
a function $\Psi_\Sigma ( j )$ of the indices $\{ j \}$ on
$\Sigma$ as
\eqn\expect{ \Psi_\Sigma( j )
              = \langle F_\Sigma( j  ; M ) \rangle
              = { 1 \over Z} \int [ dM ] F_\Sigma( j  ; M)
                           \exp(-S(M)) .}
We have not yet summed over the indices $\{ j \}$.

We saw in eq.\topological\
that the weight $I_q(T)$ for the simplicial complex $T$ in
in the summation \tensorpartition\ is given by a sum over
$\{ j \}$ on the edges of $T$. We may regard $I_q(T)$
as a partition function of a lattice statistical model
on $T$. In such a topological lattice model,
we can introduce a notion of physical states on the closed
triangulated surface $\Sigma$ as follows \viro\ \sasakura\ \ooguri .
We start with a space $V_\Sigma$ spanned by functions $\Phi( j )$
of $\{ j \}$ on $\Sigma$ and call it a space of states.
In order to define physical states in $V_\Sigma$,
we consider a three-dimensional simplicial complex $T_{\Sigma_0,\Sigma_1}$
which is homeomorphic to $\Sigma \times [0 , 1]$ and whose boundaries
$\Sigma \times \{0 \}$ and $\Sigma \times \{ 1 \}$ are triangulated as
$\Sigma_0$ and $\Sigma_1$. We then define a map $P_{\Sigma_0,\Sigma_1}$
from $V_{\Sigma_0}$ to $V_{\Sigma_1}$ by
$$ P_{\Sigma_0,\Sigma_1}:
      \Phi (j) \in V_{\Sigma_0} \rightarrow
                    \sum_{\{ j' \}} \Phi(j')
               P_{\Sigma_0,\Sigma_1}(j' , j )
                 \in V_{\Sigma_1} $$
where $P_{\Sigma_0,\Sigma_1} ( j' , j )$
is defined by attaching an index $J_e$
on each edge $e$ on $T_{\Sigma_0,\Sigma_1}$ and by summing over $\{ J \} $ as
\eqn\prop{ \eqalign{ & P_{\Sigma_0,\Sigma_1} ( j' , j )
         = {\Lambda_q^{n_0(\Sigma_1)}
               \over \Lambda_q^{n_0(T_{\Sigma_0,\Sigma_1})}}
                \sum_{\{ J \} }
         \prod_{e \in \Sigma \times \{ 0 \} }
                  \delta (J_e,j_e)\cr &~~~\times
	            \prod_{e \in \Sigma \times \{ 1 \} }
  \left( {\delta (J_e,j_e') \over (-1)^{2j_e'} [2j_e'+1]_q} \right)
       \prod_{e \in T_{\Sigma_0,\Sigma_1}} (-1)^{2J_e}[ 2J_e +1 ]_q
       \cr &~~~~~\times   \prod_{t \in T_{\Sigma_0,\Sigma_1}}
            (-1)^{\sum_{k=1}^6 J_{e_k(t)}}
  \left\{ \matrix{ J_{e_1(t)} & J_{e_2(t)} & J_{e_3(t)} \cr
                  J_{e_4(t)} & J_{e_5(t)} & J_{e_6(t)} \cr }
             \right\}_q      \cr}}
Because of the topological invariance, $P_{\Sigma_0,\Sigma_1}$ is
independent of the choice of the triangulation in the interior of
$T_{\Sigma_0, \Sigma_1}$ and satisfies \viro
\eqn\compose{ \sum_{ \{ j' \} } P_{\Sigma_0,\Sigma_1}(j,j')
                     P_{\Sigma_1,\Sigma_2}(j',j'')
                          =P_{\Sigma_0,\Sigma_2}(j,j'') }
where $\Sigma_0$, $\Sigma_1$ and $\Sigma_2$ are all homeomorphic
to $\Sigma$. Especially when $\Sigma_0=\Sigma_1=\Sigma$,
$P_{\Sigma,\Sigma}$ acts as a projection operator on $V_\Sigma$.
The physical subspace $V^{(phys)}_\Sigma$ is defined as a subspace
of $V_\Sigma$ projected out by $P_{\Sigma,\Sigma}$ as
$$ V^{(phys)}_\Sigma
     =\{ \Phi \in V_\Sigma : \sum_{\{ j' \}} \Phi(j')
      P_{\Sigma,\Sigma}(j',j) = \Phi(j) \}. $$
Because of the property \compose ,
the operator $P_{\Sigma_0,\Sigma_1}$ defines an invertible map
from a physical state in $V_{\Sigma_1}$ onto
a physical state in $V_{\Sigma_0}$.
Therefore $V^{(phys)}_{\Sigma_0}$ and
$V^{(phys)}_{\Sigma_1}$ are isomorphic.

Since $\Psi_\Sigma(j)$ given by \expect\ is also a function
of $\{ j \}$, we may consider its pairing with a physical
state $\Phi$ on $\Sigma$ as
\eqn\coupling{  ( \Phi, \Psi_\Sigma  )
     =  \Lambda_q^{{1\over 2}n_2(\Sigma)+h(\Sigma)}
            \sum_{\{ j \}} \Phi(j) \Psi_\Sigma(j)  }
where $h(\Sigma)$ is the genus of $\Sigma$.
The factor $\Lambda_q^{{1\over 2}n_2(\Sigma)+h(\Sigma)}$
will simplify the equations in the following.
As we shall see below, $\Phi_\Sigma(j)$ itself is not
a physical state, and the pairing $( \Phi , \Psi_\Sigma)$
depends on the triangulation $\Sigma$ of the surface.
The Schwinger-Dyson equation describes the dependence of
$(\Phi , \Psi_\Sigma )$ on $\Sigma$.

Among physical states on $\Sigma$, especially important are
the Hartle-Hawking type states \ooguri\ defined as follows.
A handlebody ${\cal H}_h$ of genus $h$ is
obtained by cutting out $2h$ pairs
of discs from a boundary of a three-dimensional ball $B^3$
and by identifying them pairwisely in opposite orientations.
The boundary of ${\cal H}_h$ is then a closed orientable surface of
genus $h$. The boundaries of the discs become simple homology cycles
on the boundary of ${\cal H}_h$ which do not intersect with
each other and which are contractible in ${\cal H}_h$.
They are called the meridians of the handlebody.
Let us consider a simplicial manifold ${\cal H}_\Sigma^{(\alpha)}$
homeomorphic to a handlebody of genus $h$
whose boundary is triangulated as on $\Sigma$ ($h(\Sigma)=h$)
and whose meridians are given by $\{ \alpha_1,...,\alpha_h \}$
where each $\alpha_i$ is chosen to be a sequence of edges on $\Sigma$.
The Hartle-Hawking type state $\Phi^{(\alpha)}(j)$
associated to the handlebody is defined by
\eqn\hawking{ \eqalign{
   \Phi^{(\alpha)}(j)
         = & {\Lambda_q^{n_0(\Sigma)} \over
              \Lambda_q^{n_0({\cal H}_\Sigma^{( \alpha )})}}
                \sum_{\{ J \} }
              \prod_{e \in \Sigma}
 \left( {\delta (J_e,j_e)  \over (-1)^{2j_e} [2j_e+1]_q } \right)
             ~~   \prod_{e \in {\cal H}_\Sigma^{( \alpha )}}
           (-1)^{2J_e}[ 2J_e +1 ]_q \cr \times &
         \prod_{t \in {\cal H}(\{ \alpha \})}
            (-1)^{\sum_{k=1}^6 J_{e_k(t)}}
  \left\{ \matrix{ J_{e_1(t)} & J_{e_2(t)} & J_{e_3(t)} \cr
                  J_{e_4(t)} & J_{e_5(t)} & J_{e_6(t)} \cr }
             \right\}_q      \cr}  }
Because of the topological invariance, $\Phi^{( \alpha )}(j)$
defined in this way is independent of the choice
of the triangulation of the interior
of the handlebody, and it gives a physical state on $\Sigma$.

By expanding integral \expect\ in powers of $g$,
the pairing of $\Psi_\Sigma$ with the Hartle-Hawking type state
$\Phi^{(\alpha )}$ is expressed as
\eqn\heegaard{ \eqalign{
 ( \Phi^{(\alpha)}, \Psi_\Sigma )& =
           \Lambda_q^{{1\over 2} n_2(\Sigma)+h(\Sigma)}
   \sum_{\{ j \}} \Phi^{( \alpha )}(j) \langle F(j;M) \rangle \cr
 & =      \Lambda_q^{{1\over 2} n_2(\Sigma)+h(\Sigma)}
        \sum_{T: \partial T = \Sigma}
         \left({C(\Sigma) \over C(T)}\right) (g \Lambda_q)^{n_3(T)}
 \left( {\mu \over \Lambda_q} \right)^{n_1(T)} \cr &~~~~~~~~~~~~
     ~~~~~~~~~\times
   \Lambda_q^{n_0(T)-1} I_q({\cal H}_\Sigma^{( \alpha )}
              \cup_\Sigma (-T)) \cr
& =          \sum_{T : \partial T = \Sigma}
   \left({C(\Sigma) \over C(T)}\right)    g^{n_3(T)}       \mu^{n_1(T)}
          \Lambda_q^{\chi(T)- {1\over 2} \chi(\Sigma)}
     I_q({\cal H}_\Sigma^{(\alpha )} \cup_\Sigma (-T)) \cr}}
where the sum $\sum_T$ is over simplicial
complexes whose boundary is $\Sigma$ and
$I_q({\cal H}_\Sigma^{( \alpha )} \cup_\Sigma (-T))$
is the Turaev-Viro invariant \topological\ for the simplicial complex
obtained by gluing ${\cal H}_\Sigma^{( \alpha )}$ and $T$
together on their boundaries after inverting the orientation of $T$.
As in the case of the matrix model, the dual graph of $T$ is not
necessarily connected though it does not contain a closed
disconnected part. Especially a triangle on the boundary $\Sigma$
of $T$ may be attached to another triangle on $\Sigma$.
For any simplicial complex $T$ bounded by $\Sigma$,
its Euler number $\chi(T)$ is greater than or equal to
${1\over 2}\chi(\Sigma)$, and the equality holds
if and only if ${\cal H}_\Sigma^{(\alpha)} \cup (-T)$ is a simplicial
manifold. If we could take the limit $q \rightarrow 0$, the right-hand
side in the above would become a sum over $T$ such that
$\chi(T) = {1\over 2} \chi(\Sigma)$ and
${\cal H}_\Sigma^{( \alpha )} \cup_\Sigma (-T) \simeq S^3$.
Although this limit itself is not well-defined,
we will find a set of factorization conditions which together with
the Schwinger-Dyson equation imposes the similar conditions on
$T$.

In order to derive the Schwinger-Dyson equation, we choose
a triangle $\Delta_0$ on $\Sigma$ and consider an open surface
$\Sigma_{\Delta_0}$ obtained by removing the triangle $\Delta_0$
from $\Sigma$. Associated to $\Sigma_{\Delta_0}$, we define a boundary
operator $F_{\Sigma_{\Delta_0}}(j;M)^{a_1a_2a_3}_{m_1m_2m_3}$ by
$$
 \eqalign{  F_\Sigma( j ; M)^{a_1a_2a_3}_{m_1m_2m_3}
       =  &\sum_{\{ m \}, \{ a \}}
           \prod_{r=1}^3    \delta (m_r,m_{e_r(\Delta_0)})
  \delta (a_r,a_{e_r(\Delta_0)}) \cr &\times
      \prod_{e: edges} {(-1)^{j_e} \over 2}
   \left[ (-q)^{m_e}+(-q)^{-m_e} \right]
   \cr & \times \prod_{\Delta: triangles}
             M^{j_{e_1(\Delta)}j_{e_2(\Delta)}j_{e_3(\Delta)};
              a_{e_1(\Delta)}a_{e_2(\Delta)}a_{e_3(\Delta)}}
               _{\pm m_{e_1(\Delta)}
                         \pm m_{e_2(\Delta)} \pm m_{e_3(\Delta)}}.\cr} $$
As in the case of the matrix model, we make use of the identity
$$    {1 \over Z} \int [dM]  \sum_{\{ m \} , \{ a \}}
   {\partial \over \partial
     M^{j_1 j_2 j_3 ;a_1a_2a_3}_{m_1m_2m_3}}
      \left[ F_{\Sigma_{\Delta_0}}(j;M)^{a_1a_2a_3}_{m_1m_2m_3}
                \exp(-S(M)) \right] =0 $$
where we have set $j_{e_r(\Delta_0)}=j_r$ ($r=1,2,3$). By performing
the derivative with respect to $M$, we obtain
\eqn\threedel{ \eqalign{ \Psi_\Sigma(j) =&
    g \Lambda_q \sum_{ j_4,j_5,j_6}
       (-1)^{\sum_{k=1}^6 j_k}
      \left\{  \matrix{ j_1 & j_2 & j_3 \cr
                        j_4 & j_5 & j_6 \cr } \right\}
         \Psi_{\Sigma_{\widehat{\Delta}_0}}(j)\cr
 & +\sum_{\{ m \}, \{ a \}}
    \langle  {\partial F_{\Sigma_{\Delta_0}}(j;M)^{
   a_1a_2a_3}_{m_1m_2m_3}
         \over \partial M^{j_1j_2j_3;a_1a_2a_3}_{m_1m_2m_3}}
      \rangle \cr} }
where $\Sigma_{\widehat{\Delta}_0}$ is a triangulated surface obtained
from $\Sigma$ by decomposing $\Delta_0$ into three small triangles,
and $j_4,j_5,j_6$ are spins on the three additional
edges on $\Sigma_{\widehat{\Delta}_0}$.
If it were not for the last term in the right-hand side of
eq.\threedel , $\Psi_{\Sigma}$ would transform like a physical
state under the change of triangulation $\Sigma \rightarrow
\Sigma_{\widehat{\Delta}_0}$.

Since each $M$ in $F_{\Sigma_{\Delta_0}}(j;M)$
corresponds to a triangle $\Delta$ on $\Sigma_{\Delta_0}$,
the last term $\langle {\partial \over \partial M} F_{\Sigma_{\Delta_0}}
(j;M) \rangle$ in the right-hand side of \threedel\
can be expressed a sum over closed surfaces
$\Sigma_{\Delta,\Delta_0}$ obtained by removing $\Delta$ and $\Delta_0$
from $\Sigma$ and by identifying their edges.
Let us denote the number of common edges
of the two triangles by $E(\Delta,\Delta_0)$
and the number of common vertices by $V(\Delta,\Delta_0)$.
We can classify the position of $\Delta$ on $\Sigma_{\Delta_0}$
according to the values of $E$ and $V$ as follows.

\vskip .1cm
\noindent
(1) When $E(\Delta,\Delta_0)\neq 0$ and $V(\Delta,\Delta_0)
= E(\Delta,\Delta_0)+1$, $\Delta$ is in the neighborhood of
$\Delta_0$ on $\Sigma$. In this case,
$\Sigma_{\Delta,\Delta_0}$ has the same topology as $\Sigma$.

\vskip .1cm
\noindent
(2) When $E(\Delta,\Delta_0)=0$ and $V(\Delta,\Delta_0)=0,1$,
the removal of the triangles will create a new handle on $\Sigma$
and the edges of $\Delta_0$ becomes a homology cycle around the handle.

\vskip .1cm
\noindent
(3) When $V(\Delta,\Delta_0) \geq E(\Delta,\Delta_0)+2$,
the triangles $\Delta$ and $\Delta_0$ wrap around a cycle on
$\Sigma$, and $\Sigma_{\Delta,\Delta_0}$ is
degenerate around the cycle.

\noindent
Let us consider a generic situation
when $E(\Delta,\Delta_0) + 1 \geq V(\Delta,\Delta_0)$
for any triangle $\Delta$ on $\Sigma_{\Delta_0}$, which
includes the cases (1) and (2). We also assume here
that the surface $\Sigma$ is not degenerate in
the neighborhood of $\Delta_0$. The cases involving
degenerate surfaces will be examined in the next section.

In the generic situation, by multiplying an arbitrary physical state
$\Phi$ on both sides of eq.\threedel\ and by summing over $\{ j \}$,
we obtain
\eqn\sdthree{  \eqalign{
    ( \Phi, \Psi_\Sigma ) & =
        g ( \Phi , \Psi_{\Sigma_{\widehat{\Delta}_0}} )
     + \sum_{\Delta; E(\Delta,\Delta_0) \neq 0}
          \mu^{E(\Delta,\Delta_0)}
             ( \Phi , \Psi_{\Sigma_{\Delta, \Delta_0}} ) \cr
        & ~~~~
     + \sum_{\Delta; E(\Delta,\Delta_0)=0}
         ( i[ \Phi ]_{\Delta,\Delta_0},
                 \Psi_{\Sigma_{\Delta,\Delta_0}} ). \cr} }
The second term in the right-hand side is a sum over $\Delta$
in the neighborhood of $\Delta_0$, and $\Sigma_{\Delta,\Delta_0}$
is of the same topology as $\Sigma$. On the other hand,
$h(\Sigma_{\Delta,\Delta_0})=h(\Sigma)+1$ in the last term, and
$i_{\Delta,\Delta_0}$ is defined by
$$  i[ \Phi ]_{\Delta,\Delta_0} (j) =
   \sum_{ \{ j_{e_r(\Delta)} \}_{r=1}^3 }
    \left( \prod_{r=1}^3
     \delta (j_{e_r(\Delta)},j_{e_r(\Delta_0)}) \right)
           \Phi(j). $$
It is easy to show that this gives a map from $V^{(phys)}_{\Sigma}$
into $V^{(phys)}_{\Sigma_{\Delta,\Delta_0}}$.

Especially
when $\Phi$ is the Hartle-Hawking type state $\Phi^{(\alpha)}$,
$i[\Phi^{( \alpha )}]_{\Delta,\Delta_0}$ is also
the Hartle-Hawking type state associated to a handlebody
whose boundary is $\Sigma_{\Delta,\Delta_0}$ and
whose meridians are $\{ \alpha_1,..., \alpha_{i(\Sigma)} , \partial
\Delta_0\}$
\eqn\map{  i[ \Phi^{(\alpha)} ]_{\Delta,\Delta_0}(j)
 = \Phi^{(\alpha \cup \{ \partial \Delta_0 \} )} (j) }
where $\partial \Delta_0$ is the boundary of $\Delta_0$.
The equation \sdthree\ for $\Phi=\Phi^{( \alpha )}$ can then be
written as
\eqn\schwinger{    \langle \Sigma^{(\alpha)} \rangle
      =  g \langle \Sigma_{\widehat{\Delta}_0}^{(\alpha)} \rangle
             + \sum_{\Delta; E(\Delta,\Delta_0) \neq 0}
             \mu^{E(\Delta,\Delta_0)}
      \langle \Sigma_{\Delta,\Delta_0}^{(\alpha)} \rangle
     + \sum_{\Delta; E(\Delta,\Delta_0) = 0 }
             \langle
   \Sigma_{\Delta,\Delta_0}^{(\alpha \cup \{ \partial \Delta_0 \})}
                  \rangle  }
where
\eqn\amplitude{ \langle \Sigma^{(\alpha)} \rangle
  = (\Phi^{( \alpha )}, \Psi_\Sigma )
    = \Lambda_q^{{1 \over 2}n_2(\Sigma)+h(\Sigma)}
         \sum_{\{ j \}} \Phi^{( \alpha )} (j)
         \langle F_\Sigma(j;M) \rangle. }
We call this the Schwinger-Dyson equation for the amplitude
$\langle \Sigma^{(\alpha)} \rangle$
of the triangulated surface $\Sigma$ with the choice $\{ \alpha_i \}$
of simple homology cycles. We see that, when
$\Sigma_{\Delta,\Delta_0}$ is non-degenerate,
the Schwinger-Dyson equation \schwinger\
does not depend on the value of $q$ explicitly.

\vskip 1cm

\noindent
{\bf 5. Restriction on Topologies}

\vskip .5cm

\noindent
{\bf 5-1. General Case}

\vskip .3cm

We have found that
the Schwinger-Dyson equation \schwinger\
takes the same form for any value of $q$ as far as the surface
$\Sigma$ in the equation is non-degenerate.
On the other hand, the amplitude $\langle \Sigma^{(\alpha)} \rangle$
which solves the equation is expanded as a sum over $T$  weighted with
$\Lambda_q^{\chi(T)-{1\over 2}\chi(\Sigma)} I_q
({\cal H}_\Sigma^{(\alpha)} \cup_\Sigma (-T)) $ as in eq. \heegaard .
Depending on the topology of $T$, the weight takes various
functional forms in $q$. Since the Schwinger-Dyson equation
for non-degenerate surfaces is linear in
$\langle \Sigma^{(\alpha)} \rangle$, a sum over a
restricted class of triangulations $T$ such that
the weight takes the same functional form should also give a
solution to the equation.

Here we claim a stronger statement
that, for each choice of a compact orientable three-dimensional
manifold $M$, a sum over triangulations $T$ such that
\eqn\condition{ {\cal H}_\Sigma^{(\alpha)} \cup_\Sigma (-T) \simeq M }
satisfies the Schwinger-Dyson equation. Being a triangulation of
the manifold, $T$ should also satisfy $\chi(T) = {1 \over 2}
\chi(\Sigma)$. Let us define
\eqn\restrict{ \eqalign{
  \langle \Sigma^{(\alpha)} \rangle_{|M} & =
    \sum_{T: {\cal H}_\Sigma^{(\alpha)} \cup_\Sigma (-T)
    \simeq M} \left( {C(\Sigma) \over C(T)} \right)
     g^{n_3(T)} \mu^{n_1(T)} \cr
   & = \sum_{n_3,n_1} Z_{n_3,n_1}(\Sigma;\alpha)_{|M}
                  g^{n_3} \mu^{n_1} \cr } }
where the sum $\Sigma_T$ is over triangulations $T$
satisfying the condition \condition ,
and $Z_{n_3,n_1}(\Sigma;\alpha)_{|M}$ is
the number of such triangulations with $n_3(T)=n_3$,
$n_1(T)=n_1$. We are going to show that this gives a
solution to the Schwinger-Dyson equation.

By definition, for each triangulation $T$ counted in
the above, ${\cal H}_\Sigma^{(\alpha)}
\cup_\Sigma (-T)$ give a triangulation of $M$. Let us pay
attention to one of the triangle $\Delta_0$
on the boundary $\Sigma$ of $T$. In the simplicial complex $T$,
the triangle $\Delta_0$ may be either attached to a tetrahedron
or identified with another triangle $\Delta$ on $\Sigma_{\Delta_0}$.

If $\Delta_0$ is attached to a tetrahedron in $T$,
we can remove the tetrahedron from $T$ and give it to the handlebody
${\cal H}_\Sigma^{(\alpha)}$, without changing the triangulation
of $M$. After removing the tetrahedron, $T$ becomes a
triangulation $T'$ consisting of $(n_3-1)$ tetrahedron
whose boundary is $\Sigma_{\widehat{\Delta}_0}$.
Obviously $T'$ obeys $\chi(T')
= {1\over 2}\chi(\Sigma_{\widehat{\Delta}_0})$
if $\chi(T)= {1 \over 2} \chi(\Sigma)$. The
handlebody ${\cal H}_\Sigma^{(\alpha)}$
becomes ${\cal H}_{\Sigma_{\widehat{\Delta}_0}}^{(\alpha)}$
after adding the tetrahedron on it.
Since the triangulation of $M$ remains unchanged,
$T'$ should satisfy
\eqn\conditiontwo{ {\cal H}_{\Sigma_{\widehat{\Delta}_0}}^{(\alpha)}
\cup_{\Sigma_{\widehat{\Delta}_0}} (-T') \simeq M. }
Such $T'$ is counted in $Z_{n_3-1,n_1}(\Sigma_{\widehat{\Delta}_0};
\alpha)_{|M}$. Conversely, from any triangulation $T'$ satisfying
the above, we can construct a triangulation $T$ such that
the condition \condition\ holds. However the correspondence between
$T$ and $T'$ is not necessarily one to one when $T$ and $T'$ have
non-trivial symmetries. We shall see below that this is taken care
of by the factors $C(\Sigma)/C(T)$ and $C(\Sigma_{\widehat{\Delta}_0}
)/C(T')$.

Let $G(\Sigma)$ be the symmetry of $\Sigma$ under interchange
of the triangles on it, and $G(\Sigma_{\Delta_0})$ be the
invariant subgroup of $G(\Sigma)$ which keeps the
position of the triangle $\Delta_0$ fixed.
Then $[G(\Sigma) : G(\Sigma_{\Delta_0}) ]=C(\Sigma)/C(\Sigma_{\Delta_0})$
gives the number of the triangles on $\Sigma$ which are conjugate
to $\Delta_0$ under the action of $G(\Sigma)$. Suppose that $m_1(T)$
of them are attached to tetrahedra in the interior of $T$, and
$m_2(T)$ of them are attached to triangles on $\Sigma_{\Delta_0}$
so that
$m_1(T)+m_2(T)=C(\Sigma) / C(\Sigma_{\Delta_0})$. The contribution
$C(\Sigma)/C(T)$ of the triangulation $T$ to $Z_{n_3,n_1}
(\Sigma;\alpha)_{|M}$ can then be divided into two parts as
$$ {C(\Sigma) \over C(T)} = C(\Sigma_{\Delta_0})
       \cdot \left(
    {m_1(T) \over C(T)} + {m_2(T) \over C(T)} \right). $$

Let us show that the sum
\eqn\subsum{ C(\Sigma_{\Delta_0}) \cdot
        \sum_T \left( {m_1(T) \over C(T)} \right) }
over triangulations $T$ with satisfy
eq. \condition\ and $n_3(T)=n_3$, $n_1(T)=n_1$
is equal to $Z_{n_3-1,n_1}(\Sigma_{\widehat{\Delta}_0};\alpha)_{|M}$.
If we remove a tetrahedron attached to one of the $m_1(T)$ triangles
conjugate to $\Delta_0$, we obtain a triangulation $T'$ consisting
of $(n_3-1)$ tetrahedra. It may happen that there is more
than one ways to remove a tetrahedron from $T$ to obtain the
same triangulation $T'$.
Let $N(T \rightarrow T')$ be the number of ways to obtain
$T'$ from $T$. Obviously $m_1(T)=\sum_{T'} N(T \rightarrow T')$,
and the summation given by eq. \subsum\ is equal to
$$ C(\Sigma_{\Delta_0})\cdot  \sum_T \sum_{T'} \left(
     { N(T \rightarrow T') \over C(T)} \right) .$$
The number $N(T \rightarrow T')$ can be expressed as
$$ N(T \rightarrow T') = {C(T) \over C(T,T')} $$
where $C(T, T')$ is the order of the symmetry of
$T$ which keeps the position of the tetrahedron
at $T \setminus T'$ fixed.
Therefore eq. \subsum\ can be expressed as
$$ C(\Sigma_{\Delta_0})\cdot  \sum_T \sum_{T'}
      {1 \over C(T,T')} $$
On the other hand,
$$ \eqalign{ Z_{n_3-1,n_1}(\Sigma_{\widehat{\Delta}_0}
;\alpha)_{|M} & = C(\Sigma_{\Delta_0}) \cdot
    \sum_{T'}  \left( {C(\Sigma_{\widehat{\Delta}_0})/
    C(\Sigma_{\Delta_0}) \over C(T') } \right) \cr
    & = C(\Sigma_{\Delta_0}) \cdot \sum_{T'} \sum_T \left(
         {N(T \leftarrow T') \over C(T')}  \right) \cr} $$
where $N(T \leftarrow T')$ is the number of ways to construct
$T$ from $T'$ by attaching a tetrahedron on the boundary of $T'$.
The number $N(T \leftarrow T')$ can also be expressed as
$$  N(T' \leftarrow T)
    = {C(T') \over C(T,T')} .$$
Combining these, we obtain
\eqn\dysonone{\eqalign{Z_{n_3-1,n_1}(\Sigma_{\widehat{\Delta}_0}
;\alpha)_{|M} &= C(\Sigma_{\Delta_0}) \cdot
     \sum_{T'} \sum_T {1 \over C(T,T')} \cr
    &=
C(\Sigma_{\Delta_0}) \cdot
        \sum_T \left( {m_1(T) \over c(T)} \right) .\cr}}

Next we consider the case when $\Delta_0$ on the boundary of
$T$ is identified with another triangle $\Delta$ on $\Sigma_{\Delta_0}$.
If $\Delta$ is in the
neighborhood of $\Delta_0$ on $\Sigma$, i.e. $E(\Delta,\Delta_0)
\neq 0$ and $V(\Delta,\Delta_0)
=E(\Delta,\Delta_0)+1$, we can
remove the triangle $\Delta=\Delta_0$ from $T$ to obtain
$T'$ whose boundary is $\Sigma_{\Delta,\Delta_0}$.
In this case, the topologies of the boundaries are the same,
i.e. $\chi(\Sigma)=\chi(\Sigma_{\Delta,\Delta_0})$.
Since $T'$ is obtained from $T$ by removing
one triangle, $E(\Delta,\Delta_0)$ edges and
$(V(\Delta,\Delta_0)-2)$
vertices, the Euler number of $T'$
remains the same as that of $T$;
$\chi(T')=\chi(T)-1+E(\Delta,\Delta_0)-(V(\Delta,\Delta_0)-2)
=\chi(T)$. Therefore
$\chi(T)={1\over 2} \chi(\Sigma)$ implies
$\chi(T')={1 \over 2}\chi(\Sigma_{\Delta,\Delta_0})$.
In this case,
the identification of the two triangles transforms
the handlebody ${\cal H}_{\Sigma}^{(\alpha)}$ into
${\cal H}_{\Sigma_{\Delta,\Delta_0}}^{(\alpha)}$, and
the triangles $\Delta=\Delta_0$ is
now in the interior of ${\cal H}_{\Sigma_{\Delta,
\Delta_0}}^{(\alpha)}$.
By construction, the triangulation $T'$ satisfies
${\cal H}_{\Sigma_{\Delta,\Delta_0}}^{(\alpha)} \cup_{\Sigma_{\Delta,
\Delta_0}} (-T') \simeq M$.

On the other hand, when $\Delta$ is not in the neighborhood of
$\Delta_0$, i.e. $E(\Delta,\Delta_0) = 0$ and $V(\Delta,\Delta_0)
=0,1$, the removal of the triangle $\Delta=\Delta_0$ from $T$ creates a
hole in $T$.
The Euler number of the boundary surface then decreases by two,
$\chi(\Sigma_{\Delta,\Delta_0}) = \chi(\Sigma)-2$.
On the other hand, the numbers of the edges
and the vertices of $T'$ remain the same as those of $T$. Therefore
$\chi(T') = \chi(T)-1=
{1 \over 2}\chi(\Sigma_{\Delta,\Delta_0})$.
The handlebody ${\cal H}_{\Sigma}^{(\alpha)}$
acquires a new handle after the removal of
$\Delta=\Delta_0$, and $\Delta_0$ becomes
a meridian disc of the handle. Therefore
${\cal H}_{\Sigma_{\Delta,\Delta_0}}^{(\alpha \cup \{ \partial \Delta_0
\})} \cup_{\Sigma_{\Delta,\Delta_0}} (-T') \simeq M$.

As in the case of eq. \dysonone , we can estimate
the contribution of these triangulations to $Z_{n_3,n_1}(\Sigma;
\alpha)_{|M}$, taking the symmetry factor $C(\Sigma)/C(T)$
into account, and obtain
\eqn\dysontwo{ \eqalign{
C(\Sigma_{\Delta_0}) \cdot
  \sum_T \left( {m_2(T) \over c(T)} \right)
 =&  \sum_{\Delta : E(\Delta,\Delta_0) \neq 0}
Z_{n_3,n_1-E(\Delta,\Delta_0)}
 (\Sigma_{\Delta,\Delta_0} ;\alpha)_{|M} \cr
 &+ \sum_{\Delta : E(\Delta,\Delta_0) = 0}
Z_{n_3,n_1} (\Sigma_{\Delta,\Delta_0}; \alpha \cup
 \{ \partial \Delta_0 \})_{|M}.\cr} }
Combining this with eq. \dysonone ,
$Z_{n_3,n_1}(\Sigma;\alpha)_{|M}$ is expressed as
\eqn\dyson{
 \eqalign{ Z_{n_3,n_1}(\Sigma;\alpha)
     & = Z_{n_3-1,n_1}(\Sigma_{\widehat{\Delta}_0};\alpha)
       + \sum_{\Delta: E(\Delta,\Delta_0) \neq 0}
         Z_{n_3,n_1-E(\Delta,\Delta_0)}
              (\Sigma_{\Delta,\Delta_0};\alpha) \cr
    &~~~~~~~~~
       + \sum_{\Delta: E(\Delta,\Delta_0) = 0 }
         Z_{n_3,n_1}(\Sigma_{\Delta,\Delta_0};
        \alpha \cup \{ \partial \Delta_0 \}) . \cr }}
Therefore $\langle \Sigma^{(\alpha)} \rangle_{|M}$
solves the Schwinger-Dyson equation, for any choice of
$M$.

\vskip 1cm

{\bf 5-2. Triangulations of Handlebodies}

\vskip .3cm

We have seen that the Schwinger-Dyson equation
\dyson\ holds for any closed orientable manifold $M$.
Especially when $M = S^3$, there is a nice characterization
of triangulations $T$ counted in $Z_{n_3,n_1}(\Sigma;\alpha)_{|S^3}$.

If we remove a handlebody from $S^3$, the rest is also a handlebody
which is described as follows. Let us first consider the case
when $\Sigma$ is isomorphic to a torus. In this case,
it is convenient to choose a canonical homology basis $\{ \alpha,
\beta \}$ such that $\alpha$ is the meridian of the handlebody
(solid torus)
${\cal H}_\Sigma^{(\alpha)}$ and that $\alpha$ and $\beta$ intersects
once with each other transversely.
The cycle $\beta$ is called the longitude of
the handlebody. The choice of the longitude is unique upto
$\alpha$, i.e. $\beta + n \alpha$ is also the longitude of
${\cal H}_\Sigma^{(\alpha)}$. Three-dimensional manifolds
with genus-one Heegaard splittings are classified as
$$ \eqalign{ & {\cal H}_\Sigma^{(\alpha)} \cup_\Sigma
                  (-{\cal H}_\Sigma^{(n\alpha+\beta)})
                \simeq S^3 \cr
             & {\cal H}_\Sigma^{(\alpha)} \cup_\Sigma
                  (-{\cal H}_\Sigma^{(n\alpha)})
                \simeq S^1 \times S^2 \cr
             & {\cal H}_\Sigma^{(\alpha)} \cup_\Sigma
                  (-{\cal H}_\Sigma^{(n \alpha + m \beta)})
                \simeq L_{m,n} ~~~(m \neq 0,1) \cr } $$
where $L_{m,n}$ is the lense space. Therefore,
if ${\cal H}_\Sigma^{(\alpha)} \cup_\Sigma (-T)
\simeq S^3$,
$T$ must be a triangulation of a handlebody whose meridian
(longitude) is the longitude (meridian) of ${\cal H}_\Sigma^{(\alpha)}$.

This is also the case for $h(\Sigma) \geq 2$.
Corresponding to the meridians $\{ \alpha_i \}_{i=1}^{h(\Sigma)}$
of the handlebody ${\cal H}_\Sigma^{(\alpha)}$, we choose
simple homology cycles $\{ \beta_i \}$ on $\Sigma$
such that $\alpha_i$ and $\beta_i$ ($i=1,...,h(\Sigma)$)
make a canonical homology basis, i.e. they have the following
intersection properties.
$$ \eqalign{ & \alpha_i \cdot \beta_j = \delta_{ij} \cr
  & \alpha_i \cdot \alpha_j =0~,~~\beta_i \cdot \beta_j =0 \cr} $$
As in the case of torus, the cycles $\{ \beta_i \}$ are called
the longitudes of ${\cal H}_\Sigma^{(\alpha)}$.
It is known in general that the Heegaard splitting
of $S^3$ is unique upto the stable equivalence \ref\waldhausen{
F.Waldhausen, {\sl Topology} {\bf 7} (1968) 195.}, and $T$
satisfying ${\cal H}_\Sigma^{(\alpha)} \cup_\Sigma (-T)$
should be a triangulation of a handlebody whose meridians
(longitudes) are the longitudes (meridians) of
${\cal H}_\Sigma^{(\alpha)}$.

The sum over triangulations of the handlebody given
by $\langle \Sigma^{(\alpha)} \rangle_{|S^3}$ is closely related to the
sum over triangulations of $S^3$.  Before explaining the
relation, it is instructive to examine the amplitude
$\langle \Sigma^{(\alpha)} \rangle$
given by eq.\amplitude\ and its relation
to $Z_q$. Let us consider the case when $\Sigma$
is homeomorphic to $S^2$ and is given by a single tetrahedron.
For the single tetrahedron,
the Hartle-Hawking type state $\Phi$ is given by the $6j$-symbol
and $n_2=4$. The boundary operator
$\Lambda_q^2 \sum_{j} \Phi(j) F(j;M)$ for the tetrahedron is then
equal to the quartic term in the action $S(M)$ upto the factor
${1 \over 12} g \Lambda_q$. Therefore, as in the case of the matrix
model \matrixdel , the partition function $Z_q$
and the amplitude of the tetrahedron are related as
$$ \eqalign{  \langle tetrahedron \rangle
      & = 12 \Lambda_q {1 \over Z} {\partial Z \over \partial g} \cr
      & = \sum_T \left( {12 \over C(T)} \right)
   n_3(T) g^{n_3(T)-1} \mu^{n_1(T)}
                \Lambda_q^{\chi(T)} I_q(T)
                .\cr} $$
On the other hand,
$\langle tetrahedron \rangle$ is also
expanded as a sum over simplicial complexes $T$ bounded by
the tetrahedron as in eq. \heegaard . Thus
the number of closed simplicial complexes $T$ weighted with
$\Lambda_q^{\chi(T)} I_q(T)$ and the symmetry factor $1/C(T)$
is equal to $12 n_3$ times the number of open simplicial
complexes $T$ such that $\partial T = (tetrahedron)$
weighted with $\Lambda_q^{\chi(T)-2} I_q((tetrahedron)
\cup (-T))$ and the symmetry factor $(12/C(T))$.

The similar relation holds between
$\langle \Sigma^{(\alpha)} \rangle_{|S^3}$ and
the sum over triangulations of $S^3$. When $\Sigma$ is
a single tetrahedron, the amplitude is a sum over triangulations
of a three-dimensional ball $B^3$ whose boundary is the tetrahedron.
$$  \langle  tetrahedron  \rangle_{|S^3}
    = \sum_{n_3,n_1} Z_{n_3,n_1}(tetrahedron)_{|S^3}
      \cdot g^{n_3} \mu^{n_1}  $$
Here $Z_{n_3,n_1}(tetrahedron)_{|S^3}$ is the number of triangulations
of the interior of the tetrahedron
with $n_3$ tetrahedra and $n_1$ edges weighted with the
symmetry factor $(12/C(T))$. The factor $12$ is the order of
the symmetry of the tetrahedron.  We can show that
this is related to $Z_{n_3,n_1}(S^3)$ as
\eqn\sphereball{ Z_{n_3-1,n_1}(tetrahedron)_{|S^3} =
                             12 n_3 Z_{n_3,n_1}(S^3). }
Any triangulation of $B^3$ is obtained by removing
a tetrahedron from a triangulation of $S^3$, and conversely
any triangulation of $S^3$ is given by gluing a tetrahedron
to a triangulation of $B^3$. For each triangulation $T$
of $S^3$,  there are $n_3(T)$
distinct triangulations of $B^3$ as far as $T$ does not have
a symmetry. On the other hand,
if $T$ has a non-trivial symmetry, different choices
of tetrahedra in $T \simeq S^3$ related by the symmetry
give the same triangulation of $B^3$.
The symmetry factors in $Z_{n_3,n_1}(tetrahedron)$
and $Z_{n_3,n_1}(S^3)$ compensate for the
overcounting due to the symmetries,
and eq. \sphereball\ holds.

\vskip 1cm

\noindent
{\bf 6. Factorization}

\vskip .3cm

In the last section, we have shown that
$\langle \Sigma^{(\alpha)} \rangle_{|S^3}$
gives a solution to the Schwinger-Dyson equation when
$E(\Delta,\Delta_0)+1 \geq V(\Delta,\Delta_0)$ for any
$\Delta$ on $\Sigma$ and when the
surface $\Sigma$ in the equation is not degenerate.
In this section, we will define a set of factorization
conditions on $\langle \Sigma^{(\alpha)} \rangle_{|S^3}$
so that the Schwinger-Dyson equation becomes applicable even
when it involves degenerate surfaces. We will then show that
the Schwinger-Dyson equation combined with the factorization
conditions has a unique solution. Thus they characterize
the sum over triangulations $T$ such that
${\cal H}_\Sigma^{(\alpha)} \cup_\Sigma (-T) \simeq S^3$.

Let us first consider a case when the surface $\Sigma$
is degenerate.
When the pinching cycle is homologically trivial on
$\Sigma$, the surface is actually separated into two
components $\Sigma_1$ and $\Sigma_2$ touching to each
other at the vertex. In this case, $T$ must also consists
of two disconnected components $T_1$ and $T_2$ such that
$\partial T_i = \Sigma_i$ ($i=1,2$). The Schwinger-Dyson
equation remains applicable in this case if we supplement
a condition that $\Delta$ in the summation in the
right-hand side of eq. \dyson\ is restricted
to be those on the same boundary component as $\Delta_0$ does.
Especially when
we can choose the meridians of the handlebody
${\cal H}^{(\alpha)}_\Sigma$ such that
\eqn\separation{\{ \alpha_1,...,\alpha_{h(\Sigma)} \} =
    \{ \alpha_1^{(1)}, ..., \alpha_{h(\Sigma_1)}^{(1)} \} \cup
    \{ \alpha_1^{(2)}, ..., \alpha_{h(\Sigma_2)}^{(2)} \} }
where
$\alpha^{(1)}_i \in \Sigma_1$ and $\alpha^{(2)}_i \in \Sigma_2$,
the amplitude $\langle \Sigma^{(\alpha)} \rangle_{|S^3}$
factorizes as
$$ \langle \Sigma^{(\alpha)} \rangle_{|S^3}
    = \langle \Sigma^{(\alpha^{(1)})}_1 \rangle_{|S^3}
      \langle \Sigma^{(\alpha^{(2)})}_2 \rangle_{|S^3} $$
In this case, it is obvious that
$\Delta \in \Sigma_i$ should be imposed in eq. \dyson\
when $\Delta_0 \in \Sigma_i$ ($i=1,2$).

It is instructive to compare this behavior of the amplitude
with that of $\langle \Sigma^{(\alpha)} \rangle
= (\Phi^{(\alpha)},\Psi_\Sigma)$.
When $\Delta$ and $\Delta_0$ wrap around the homologically trivial
cycle $\gamma$ and when the longitudes $\{ \alpha_i \}$
are separated as in eq. \separation , $i_{\Delta,\Delta_0}$ defined by
\map\ gives a map from $V_{\Sigma}^{(phys)}$
into $V_{\Sigma_1}^{(phys)} \otimes V_{\Sigma_2}^{(phys)}$ as
$$ 	i[ \Phi^{(\alpha)} ]_{\Delta,\Delta_0}
     =  \Phi^{(\alpha^{(1)})} \Phi^{(\alpha^{(2)})} .  $$
The contribution of the degenerate surface
$\Sigma_{\Delta,\Delta_0}$ to the Schwinger-Dyson equation is
then
$$ \eqalign{ & \langle \Sigma_{\Delta,\Delta_0}^{(\alpha)} \rangle \cr &
  = \langle \Sigma_1^{(\alpha_1)} \Sigma_2^{(\alpha_2)} \rangle \cr
 &= \Lambda_q^{{1 \over 2} n_2(\Sigma_1)+h(\Sigma_1)}
    \Lambda_q^{{1 \over 2} n_2(\Sigma_2)+h(\Sigma_2)} \cr
         &~~~~~~~~~~\times
    \sum_{\{ j^{(1)} \},\{ j^{(2)} \} } \Phi^{(\alpha^{(1)})}(j^{(1)})
     \Phi^{(\alpha^{(2)})}(j^{(2)})
       \Psi_{\Sigma_1 +\Sigma_2}(j^{(1)},j^{(2)}) \cr
 &= \langle \Sigma_1^{(\alpha^{(1)})}\rangle \langle
               \Sigma_2^{(\alpha^{(2)})} \rangle
+  \Lambda_q \cdot \sum_{T: \partial T = \Sigma_1 + \Sigma_2}
    \left( {C(\Sigma_1) C(\Sigma_2) \over C(T)} \right)
   g^{n_3(T)} \mu^{n_1(T)} \cr
 &~~~~~~~~~~~~~~~~~~~~~~~~~~\times
 \Lambda_q^{\chi(T)-{1\over 2} \chi(\Sigma_1)-{1 \over 2} \chi(\Sigma_2)}
    I_q({\cal H}_{\Sigma_1}^{(\alpha^{(1)})} \cup_{\Sigma_1} (-T)
          \cup_{\Sigma_2} {\cal H}_{\Sigma_2}^{(\alpha^{(2)})} )
    .\cr }$$
In the first term in the right-hand side,
the sum is over simplicial complexes
consisting of two disconnected components
$T_1$ and $T_2$ each of which is attached to $\Sigma_1$ and
$\Sigma_2$ respectively, and the second term is a sum over
connected simplicial complexes. We see that the second term
is multiplied by the extra-factor of $\Lambda_q$. This is
interpreted as follows. If $T$ remains connected after pinching around
the cycle $\gamma$, $T$ should contain a homotopy cycle
linking with $\gamma$.  The extra-factor $\Lambda_q$ is associated with
this homotopy cycle. The factorization condition may be regarded as
the $q \rightarrow 0$ limit of this equation for $\langle
\Sigma^{(\alpha)} \rangle$ where the second term is to be suppressed
by the factor of $\Lambda_q$.

When the surface $\Sigma$ is degenerate around a homologically
non-trivial cycle, the sum over $\Delta$ in the Schwinger-Dyson
equation runs over all triangles on $\Sigma_{\Delta_0}$ as far as
$\Delta_0$ is not on the degeneration point. When one of
the vertices of $\Delta_0$ is on the degeneration point,
we must supplement the following rule: The sum over triangles
in the Schwinger-Dyson equation should not include those
in the neighborhood of $\Delta_0$ and on the other side
of the degeneration point across the degenerating cycle.
The reason for this is the following.
If we take a triangle $\Delta'$ which shares the degeneration
point with $\Delta_0$ and which is on the other side of
the point, the identification of $\Delta'$ and $\Delta_0$
does not change the Euler number of $\Sigma$. The number of
triangles in $T$, on the other hand, decreases by one
if $\Delta'$ and $\Delta_0$ were identified in $T$ and
removed from $T$. Thus the equality
$\chi(T) = {1 \over 2}\chi(\Sigma)$ cannot be sustained.
On the other hand, if we take $\Delta$ to be other than
those triangles, there is a triangulation $T$ such that
${\cal H}_\Sigma^{(\alpha)} \cup (-T) \simeq S^3$
and $\Delta$ and $\Delta_0$ are identified. Thus the
Schwinger-Dyson equation holds with this supplementary condition.

So far, we have examined the situation when
$\Sigma$ is already degenerate. When
$V(\Delta,\Delta_0) \geq E(\Delta,\Delta_0)+2$,
the triangles $\Delta$ and $\Delta_0$ wrap
around a cycle $\gamma$ on $\Sigma$ and the surface
will develop additional degeneration if we identify
$\Delta$ and $\Delta_0$ and remove them. If it is possible to construct
a triangulation $T$ in which the two
triangles $\Delta$ and $\Delta_0$ are attached to each other,
$T$ should be counted in $\langle \Sigma^{(\alpha)} \rangle_{|S^3}$.
The Schwinger-Dyson equation \schwinger\ derived for non-degenerate
surfaces becomes applicable in this case if we take into account
the contribution of the degenerate surface $\Sigma_{\Delta,\Delta_0}$.
On the other hand, if there is no such triangulation in which
$\Delta$ is identified with $\Delta_0$,
the surface $\Sigma_{\Delta,\Delta_0}$ should not be included
the Schwinger-Dyson equation.

When the cycle $\gamma$ is homologically trivial,
it is always possible to construct a triangulation $T$
in which $\Delta$ and $\Delta_0$ are identified.
Thus the term $\langle \Sigma_{\Delta,\Delta_0}^{(\alpha)}
\rangle$ should be included in the Schwinger-Dyson equation.
In this case, the surface $\Sigma_{\Delta,\Delta_0}$
is pinched around at the common vertex of $\Delta$ and
$\Delta_0$, and the Schwinger-Dyson equation can be iterated
with the supplementary conditions introduced in the above.

On the other hand,
when the cycle $\gamma$ is homologically non-trivial on $\Sigma$,
the triangles $\Delta$ and $\Delta_0$ wrap around
a handle on $\Sigma$. In this case, the handle is pinched
on the surface $\Sigma_{\Delta,\Delta_0}$.
If there is a triangulation $T$
in which the triangles $\Delta$ and $\Delta_0$ are identified with
each other, there must be a disc in $T$ whose boundary is
the cycle $\gamma$.
After removing the triangle $\Delta=\Delta_0$ from $T$, we obtain
$T'$ which gives a triangulation of a handlebody whose
boundary is $\Sigma_{\Delta,\Delta_0}$. The surface
$\Sigma_{\Delta,\Delta_0}$ is pinched at the common vertex
of $\Delta$ and $\Delta_0$. The complement of $T'$ in
$S^3$ should also be a handlebody
${\cal H}_{\Sigma_{\Delta,\Delta_0}}^{(\alpha)}$
which is pinched around the cycle $\gamma$.
The process ${\cal H}_\Sigma^{(\alpha)} \cup (-T)
\rightarrow {\cal H}_{\Sigma_{\Delta,\Delta_0}}^{(\alpha)}
\cup (-T')$ then gives the reduction
of the Heegaard splitting of $S^3$ which is the inverse
of the stabilization \waldhausen. For this to be possible,
there must be a disc in the handlebody ${\cal H}_\Sigma^{(\alpha)}$
whose boundary is on $\Sigma$ and intersects once with
$\gamma$ transversely.
If this is the case, we can choose the meridians of
${\cal H}_\Sigma^{(\alpha)}$ so that $\alpha_1 \cdot \gamma=1$
and $\alpha_i \cdot \gamma = 0$ ($i=2,...,h(\Sigma)$).
The interior of ${\cal H}_{\Sigma_{\Delta,\Delta_0}}^{(\alpha_1,...,
\alpha_{h(\Sigma)})}$ is then homeomorphic to
that of ${\cal H}_{\Sigma_{\Delta,\Delta_0}'}^{(\alpha_2,...,\alpha_{
h(\Sigma)})}$, where $\Sigma_{\Delta,\Delta_0}'$ is
obtained from $\Sigma_{\Delta,\Delta_0}$ by splitting
the common vertex of $\Delta$ and $\Delta_0$ into two
so that the pinched handle is removed;
$h(\Sigma_{\Delta,\Delta_0}')=h(\Sigma_{\Delta,\Delta'})-1$.
In this case, a term $\langle \Sigma_{\Delta,\Delta_0}
'^{(\alpha)} \rangle_M$ must be included in
the Schwinger-Dyson equation.
%
%
Conversely, if it is not possible to find such a disc in
${\cal H}_\Sigma^{(\alpha)}$, there is no
triangulation in which $\Delta$ and $\Delta_0$ are identified with each other.
In this case, there should be no term in the Schwinger-Dyson equation
corresponding to the surface $\Sigma_{\Delta,\Delta_0}$.

%
%
%
%
%
%
%

With these supplementary conditions, we can apply
the Schwinger-Dyson equation even when $\Sigma$ is
degenerate or when there is a triangle $\Delta$
such that $V(\Delta,\Delta_0) \geq E(\Delta,\Delta_0)
+2$. As in the case of the matrix model, we can show that
the Schwinger-Dyson equation combined with these
conditions can be solved inductively. The Schwinger-Dyson
equation expressed as in eq.\dyson\ is a recurrence relation for
$Z_{n_3,n_1}(\alpha;\Sigma)$ along the bibliographical ordering:
$(n_3,\Sigma) \succ (n_3',\Sigma')$ if $n_3 > n_3'$
or $n_3=n_3'$ and $n_2(\Sigma) > n_2(\Sigma')$. Because of
the factorization conditions, we can trace back the Schwinger-Dyson
equation to the expectation value of the null surface
$\langle 1 \rangle$ which is normalized to be one.
Thus the Schwinger-Dyson equation combined with the
factorization conditions has a unique solution. The solution
gives the sum over triangulations of handlebodies,
and it is related to $Z_{n_3,n_1}(S^3)$ as in eq. \sphereball .

\vskip 1cm

\centerline{{\bf Acknowledgement}}

I would like to thank Jan Ambj\o rn, Bergfinnur Durhuus,
Tohru Eguchi, Antal Jevicki and Takao Matsumoto
for discussions and comments. I would also like to thank
Michael Atiyah and Peter Goddard for their hospitality
at
the Isaac Newton Institute for Mathematical Sciences,
University of Cambridge,
where part of this work was done.
This research is supported in part by Grand-in-Aid
for Scientific Research on Priority Areas 231 "Infinite Analysis"
from the Ministry of Education, Science and Culture of Japan.

\listrefs

\end

\vskip 1cm

{\bf 7. Discussion}

\vskip .3cm
We have define the Schwinger-Dyson equation and the factorization
conditions which characterize the amplitudes of the boundary
operators in the simplicial quantum gravity in three dimensions.

In the matrix model for the two-dimensional simplicial gravity,
it is known that there is a close connection between
the Schwinger-Dyson equation and the Wheeler-de Witt equation.
In ref. \moore, it is shown that the two-point function
$\psi(L) = \langle {\sl tr}(M^{L/a}) O(M) \rangle$
with an appropriate choice of a local operator $O(M)$
gives a solution to the Bessel equation in the continuum
limit if we take $a \rightarrow 0$ simultaneously.
If we identify $\psi(L)$
as a wave-function of the Liouville model with $L$ being
the mini-superspace coordinate, the Bessel equation can be
regarded as the Wheeler-de Witt equation of the model
\ref\dhoker{E.D'Hoker and R.Jackiw, {\sl Phys. Rev Lett.}
{\bf 50} (1983) 1719; {\sl Phys. Rev.} {\bf D26}
(1982) 3517; E.D'Hoker, D.Freedman and R.Jackiw,
{\sl Phys. Rev} {\bf D28} (1983) 2583.}

The two-point function $\psi(L)$ also solves the
Schwinger-Dyson equation, which is derived from
$$ 0 = \int [dM] {\partial \over \partial M_{ij}}
  \left[ (M^{n-1})_{ij} O(M) e^{-S(M)} \right]. $$
The Schwinger-Dyson equation becomes $\psi(L)$
in the large-$N$ limit upto a contact term of
${\sl tr}(M^n)$ and $O(M)$:
\eqn\twopointsd{ \psi(L) = g \psi(L + a) + 2 \sum_k u_{k-2} \psi(L-ka)
   + \langle (M^{n-1})_{ij} {\partial \over \partial M_{ij}}
               O(M) \rangle .}
The linear part of the equation has the same form
as the linearization of the Schwinger-Dyson equation
for the one-point function $u_n = \langle
{\sl tr}(M^n) \rangle$ around its classical solution:
$u_n \rightarrow u_n + \psi(an)$, and one can
regard this as the discretized Wheeler-de Witt
operator acting on $\psi(L)$.
%
%
%
%
%
%
%
Although
the Schwinger-Dyson equation for $u_n$ is non-linear,
the equation for $\psi(L)$ is almost linear and
reduces to the Bessel equation in the continuum
limit \ref\dasjevicki{S.R.Das and A.Jevicki,
{\sl Mod. Phys. Lett.} {\bf A5} (1990) 1639},
\ref\jevicki{A.Jevicki, in {\sl }.}.
The linearized
equation contains a convolution of the two-point
function $\psi(L)$ and the one-point function $u_n$,
which reduces to a second order differential of
$\psi(L)$ in $L$ in the continuum limit.

The situation is different in the three-dimensional case
studied in this paper. There the Schwinger-Dyson equation
for the one-point function
$\langle \Sigma^{(\alpha)} \rangle_{|S^3}$ is already
linearized as far as the surface $\Sigma$ is non-degenerate.
The discretized Wheeler-de Witt equation then takes
the same form as the Schwinger-Dyson equation for
$\langle \Sigma^{(\alpha)} \rangle_{|S^3}$.
The equation, however, contains surfaces
of different topologies; e.g. $h(\Sigma_{\Delta,\Delta_0})
=h(\Sigma)+1$ if $E(\Delta,\Delta_0) =0$.
This is due to the fact that the three-dimensional
manifold in the interior of the boundary surface $\Sigma$ may
have a degenerate metric with respect to which
parts of $\Sigma$ are attached to each other.
It should be stressed that this phenomenon takes
place even if we impose the condition
$H^{(\alpha)} \cup_\Sigma (-T) \simeq S^3$ for
the topology of $T$. In the present formulation of
the model, it seems to be inevitable that such terms
appear in the regularized Wheeler-de Witt equation.
It is important  to examine if they survive
in the continuum limit of the model.

\noindent
{\bf 7. Asymptotic Solutions}

\vskip .3cm
In this section, we examine asymptotic behavior of the
solution $\langle \Sigma^{(\alpha)} \rangle_{S^3}$ to
the Schwinger-Dyson equation, assuming that
$Z_{n_3}(S^3)=\sum_{n_1} Z_{n_3,n_1}(S^3)$ grows
exponentially in $n_3$. The exponential growth
of $Z_{n_3}(S^3)$ has been suggested by
the numerical simulations \ambjorn\ \migdal\
\denmark . We have seen in Section 5
that the number of triangulations of $S^3$ is related to
the number of triangulations of a tetrahedron
$Z_{n_3,n_1}(tetrahedron)_{S^3}$ as eq. \sphereball.
Therefore $Z_{n_3,n_1}(S^3)$ itself is in principle
determined by solving the Schwinger-Dyson equation
although we have not been able to do so yet.

As in the case of the two-dimensional simplicial gravity
\ref\durhuus{B. Durhuus},
we can show that, for any $(\Sigma,\alpha)$, there
are integers $n_i(\Sigma,\alpha), n'_i(\Sigma,\alpha)$
($i=1,3$) such that
\eqn\ineq{ Z_{n_3-n_3(\Sigma,\alpha),n_1-n_1(\Sigma,\alpha)}
   (tetrahedron)
   \leq Z_{n_3,n_1}(\Sigma,\alpha) \leq
   Z_{n_3+n_3'(\Sigma,\alpha),n_1+n_1'(\Sigma,\alpha)}
   (tetrahedron). }
This inequality can be shown by noting that
any handlebody can be embedded in $B^3$ and vice versa.

The inequality \ineq\ implies the following.
If $Z_{n_3}(S^3)$ grows exponentially in $n_3$
\eqn\expgrowth{ Z_{n_3}(S^3) \simeq n_3^{-\gamma-2} e^{\beta_cn_3}
   }
for $n_3 \rightarrow \infty$,
the inequality implies that $Z_{n_3}(\Sigma;\alpha)$ also behaves as
\eqn\decompexp{    Z_{n_3}(\Sigma;\alpha) \simeq
   \psi(\Sigma;\alpha) \cdot n_3^{-\gamma-1}e^{\beta_c n_3} }
where $\psi(\Sigma;\alpha)$ does not depend on $n_3$.
On the other hand, if $Z_{n_3}(S^3)$ grows factorially in $n_3$
$$ Z_{n_3}(S^3) \simeq e^{\kappa n_3 \log n_3}, $$
the asymptotic behavior of $Z_{n_3}(\Sigma;\alpha)$ should be
\eqn\decompfact{ Z_{n_3}(\Sigma;\alpha) \simeq n_3^{\gamma(\Sigma;\alpha)}
                   \psi(\Sigma;\alpha) e^{\kappa n_3 \log n_3}. }
Here the exponent $\gamma$ may depend on $\Sigma$ and $\alpha$.

Let us first examine the case when $Z_{n_3}(S^3)$ grows exponentially.
In this case, $\langle \Sigma^{(\alpha)} \rangle_{S^3}$ is
convergent for $g \leq e^{-\beta_c}$.
By substituting \decompexp\ into the Schwinger-Dyson equation
\dyson\ and taking the limit $n_3 \rightarrow \infty$ , we obtain
\eqn\asymptdyson{ \eqalign{ \psi(\Sigma;\alpha)
     & = e^{-\beta_c}
        \psi(\Sigma_{\widehat{\Delta}_0};\alpha)
       + \sum_{\Delta: E(\Delta,\Delta_0) \neq 0}
           \psi(\Sigma_{\Delta,\Delta_0};\alpha) \cr
    &~~~~~~~~~
       + \sum_{\Delta: E(\Delta,\Delta_0) = 0 }
         \psi( \Sigma_{\Delta,\Delta_0};\alpha \cup
                \{ \partial \Delta_0 \}) .\cr} }
On the other hand, one of the factorization conditions becomes
\eqn\asymptfact{ \psi(\Sigma_1+ \Sigma_2;\alpha)
   = \psi(\Sigma_1) \langle \Sigma_2^{(\alpha_2)} \rangle_{S^3}^{(\beta_c)}
    +\psi(\Sigma_2) \langle \Sigma_1^{(\alpha_1)} \rangle_{S^3}^{(\beta_c)}}
for $\gamma > 0$. Here $\langle \Sigma^{(\alpha)} \rangle_{S^3}^{(\beta_c)}$
is the value of the amplitude at $g=e^{-\beta_c}$ and $\mu=1$,
which is convergent in this case. For $\gamma \leq 0$, the same
equation becomes
$$ \psi(\Sigma_1+\Sigma_2;\alpha) = n_3^{-\gamma}
   {\Gamma(-\gamma)^2 \over \Gamma(-2 \gamma)} \psi(\Sigma_1;\alpha)
                   \psi(\Sigma_2;\alpha_2) $$
and does not make sense in the limit $n_3 \rightarrow \infty$.

When $\gamma$ is positive, the equations \asymptfact\
and \asymptdyson\ for $\psi(\Sigma;\alpha)$
take the same forms as the Schwinger-Dyson equation \schwinger\
and the factorization condition for the
amplitude $\langle \Sigma^{(\alpha)} \rangle$ at $g=e^{-\beta_c}$
and $\mu=1$. If $\gamma$ is positive, $\langle \Sigma^{(\alpha)}
\rangle$ is convergent at these values of $g$ and $\mu$.
Similarly, we can show that the equations \asymptfact\
and \asymptdyson\ has a non-zero solution, although the
solution is not unique. Thus the ansatz
$$ Z_{n_3}(\Sigma;\alpha) \simeq
       \psi(\Sigma;\alpha) n_3^{-\gamma-1} e^{\beta_c n_3} $$
is consistent with the Schwinger-Dyson equation when $\gamma>0$.

Because of the relation \sphereball,  $Z_{n_3}(S^3)$ becomes
$n_3^{-\gamma-2} e^{\beta_c n_3}$ in this case.
The critical behavior
of the partition function $Z_{d=3}$ of the simplicial gravity
near $\beta = \beta_c$ ($g=e^{-\beta}$) is then given by
$$  Z_{d=3} \sim (\beta-\beta_c)^{\gamma+1}+({\sl regular})~~~
       (\beta \rightarrow \beta_c+0, ~~\mu=1) .$$
We have not been able to determine the explicit value of
the critical exponent $\gamma$ from this asymptotic analysis.

On the other hand, if we assume that $f(n_3)$ behaves as
$(n_3!)^\kappa$ for
$n_3 \rightarrow \infty$, the factor $c(n_3)$ becomes
$$ c(n_3) = {1 \over f(n_3)} \sum_m f(m) f(n_3-m)
       \sim 2^{-\kappa n_3} ~~~~~(n_3 \rightarrow \infty). $$
The factorization condition \asymptfact\ is then
$$ \psi(\Sigma_1+\Sigma_2;\alpha)=0 .$$
In this case, the Schwinger-Dyson equation gives
$$ \psi(\Sigma;\alpha) = \sum_{Delta:E(\Delta,\Delta_0) \neq 0}
                          \psi(\Sigma_{\Delta,\Delta_0};\alpha)
             + \sum_{\Delta:E(\Delta,\Delta_0)=0}
                \psi(\Sigma_{\Delta,\Delta_0}:
                      \alpha \cup \{ \partial \Delta_0 \} ) .$$
Since the amplitude $\langle 1 \rangle_{S^3}$ is normalized to be $1$,
$\psi$ for the null surface must be zero. With this initial condition,
the Schwinger-Dyson equation combined with the factorization conditions
implies $\psi(\Sigma;\alpha)=0$ for any $\Sigma$ and $\alpha$.
Thus there is no non-trivial solution which behaves as
$Z_{n_3}(\Sigma;\alpha) \simeq \psi(\Sigma;\alpha) (n_3!)^\kappa$.